\documentclass[twocolumn,traditabstract]{aa}
\usepackage{graphicx,amsmath}
\usepackage{natbib}
\usepackage{stfloats}
\usepackage{subfigure}
\usepackage{multirow}
\usepackage{color}
\usepackage{txfonts}
\usepackage{upgreek}
\usepackage{url}
\usepackage{epsfig}
\usepackage{epsf}
\usepackage{float}
\usepackage{slashbox}
\usepackage{sidecap}
\usepackage{fancyhdr}
\usepackage{afterpage}
\usepackage{soul}

\newcommand{\dpsiname}{polarization angle dispersion function}

\newcommand{\qq}{\sigma^2_{\rm Q}}
\newcommand{\uu}{\sigma^2_{\rm U}}

\newcommand{\ang}{\psi}

\newcommand{\snrpo}{p_0/\sigma_p}
\newcommand{\sigp}{\sigma_p}

\newcommand{\epseff}{\varepsilon_{\rm{eff}}}

\newcommand{\randval}{\pi/\sqrt{12}}

\newcommand{\mx}{\mathbf{x}}
\newcommand{\mxl}{\mathbf{x}+\mathbf{l}_i} 
\newcommand{\dpsi}{\mathcal{S}}
\newcommand{\dpsiz}{\mathcal{S}_{0}}
\newcommand{\angz}{\psi_{0}}
\newcommand{\eps}{\varepsilon}

\newcommand{\cl}{\hat{\dpsi_{C}}}

\newcommand{\dich}{\hat{\dpsi_{D}}}
\newcommand{\biascl}{Bias_{\dpsiz}}
\newcommand{\biasang}{Bias_{\dpsiz,\angz}}
\newcommand{\biassig}{Bias_{\dpsiz, \, \Sigma}}
\newcommand{\biasjoint}{Bias_{\dpsiz, \, \Sigma, \, \angz}}
\newcommand{\sigcl}{\sigma_{\dpsi,C}}
\newcommand{\sbar}{\bar{\mathcal{S}_{0}}}
\newcommand{\sn}{{S/N}}

\newcommand{\poly}{\hat{\dpsi_{P}}}

\newcommand{\newQ}{\widetilde{Q}}
\newcommand{\newU}{\widetilde{U}}
\newcommand{\newI}{\widetilde{I}}

\newcommand{\angstructname}{polarization angle structure function}

\newcommand{\low}{\textit{low}}
\newcommand{\canonical}{\textit{canonical}}
\newcommand{\extreme}{\textit{extreme}}

\title{Polarization measurement analysis}
\subtitle{III. Analysis of the {\dpsiname} with high precision polarization data}
\authorrunning{D. Alina}
\titlerunning{Polarization Measurement Analysis. III.}
\author{D. Alina\inst{1,3}, L. Montier\inst{2,3}, I. Ristorcelli\inst{2,3}, J.-P. Bernard\inst{2,3}, F. Levrier\inst{4}, E. Abdikamalov\inst{1}}
\institute{ 
\inst{1} Department  of  Physics,  School  of  Science  and  Technology,
Nazarbayev University, Astana 010000, Kazakhstan\\
\inst{2} Universit\'{e} de Toulouse, UPS-OMP, IRAP, F-31028 Toulouse cedex 4, France\\
\inst{3} CNRS, IRAP, 9 Av. colonel Roche, BP 44346, F-31028 Toulouse cedex 4, France\\
\inst{4} LERMA/LRA - ENS Paris et Observatoire de Paris, 24 rue Lhormond, 75231 Paris Cedex 05, France
}

\begin{document}

\abstract{
The high precision polarization measurements, such as those from the \textit{Planck} satellite, open new opportunities for the study of the magnetic field structure as traced by polarimetric measurements of the interstellar dust emission.
The polarization parameters suffer from bias in the presence of measurement noise.
It is critical to take into account all the information available in the data in order to accurately derive these parameters.
In our previous work, we studied the bias on polarization fraction and angle, various estimators of these quantities, and their associated uncertainties.
The goal of this paper is to characterize the bias on the {\dpsiname} that is used to study the spatial coherence of the polarization angle.
We characterize for the first time the bias on the conventional estimator of the {\dpsiname} and show that it can be positive or negative depending on the true value.
Monte Carlo simulations are performed in order to explore the impact of the noise properties of the polarization data, as well as the impact of the distribution of the true polarization angles on the bias.
We show that in the case where the ellipticity of the noise in $(Q, \, U)$ varies by less than $10 \, \%$, one can use simplified, diagonal approximation of the noise covariance matrix.
In other cases, the shape of the noise covariance matrix should be taken into account in the estimation of the {\dpsiname}.
We also study new estimators such as the dichotomic and the polynomial estimators. 
Though the dichotomic estimator cannot be directly used to estimate the {\dpsiname}, we show that, on the one hand, it can serve as an indicator of the accuracy of the conventional estimator and, on the other hand, it can be used for deriving the polynomial estimator. 
We propose a method for determining the upper limit of the bias on the conventional estimator of the {\dpsiname}. 
The method is applicable to any linear polarization data set for which the noise covariance matrices are known.
}


\maketitle

\section{Introduction}
\label{sec_introduction}

The linear polarization of the incoming radiation can be described by the Stokes parameters $Q$ and $U$ along with the total intensity $I$.
The polarization fraction $p$ and the polarization angle $\ang$ are derived from $I$, $Q$ and $U$, and bias on these parameters appears in the presence of measurement noise \citep{serkowski1958,wardle1974,simmons1985,vaillancourt2006,quinn2012}.
This issue has recently been addressed by \cite{Montier1,Montier2}, hereafter Papers I and II of this series on the polarization measurement analysis of high precision data. 
In this work, which we refer to as Paper III, we aim to characterize the bias on the {\dpsiname} - a polarization parameter that measures the spatial coherence of the polarization angle.

The interstellar magnetic field structure can be revealed by the polarimetric measurements of synchrotron radiation and of dust thermal emission and extinction \citep{mathewson1970,han2002,beck2004,heiles2005,fletcher2010}.
The interstellar dust particles are aligned with respect to the magnetic field \citep{hall1949,hiltner1949,lazarian2008}.
This leads to linear polarization in the visible, infrared and submillimetre \citep{benoit2004,vaillancourt2007,andersson2015}.
The interstellar dust polarization yields information about the direction of the plane-of-the-sky (POS) component of the magnetic field. 
\cite{heiles1996} used observations of polarization by dust extinction and found that the inclination of the Galactic magnetic field with respect to the plane of the disk of matter is about $7^{\circ}$. 
\cite{planck2014-xix} derived the all-sky magnetic field direction map as projected onto the POS from the \textit{Planck} Satellite data. 
They also used the {\dpsiname} and studied its correlation with the polarization fraction. 
In the framework of their analysis, the observed anti-correlation allowed to come to a conclusion that the observed polarization at large scales (diffuse ISM, large molecular clouds) largely depends on the magnetic field structure.
Polarimetric measurement of the emission from molecular clouds and star forming regions help to better understand the role of the magnetic field in star formation \citep{matthews2009,dotson2010,tang2012,zhang2014,cortes2016}.

\cite{DG51} and \cite{Chandrasekhar&Fermi1953} calculated the angular dispersion in polarimetric measurements of distant stars \citep{hiltner1951} to derive the strength of the magnetic field in the local spiral arm. 
Since then, the so-called Davis-Chandrasekhar-Fermi method has been widely used to derive some properties of the magnetic field such as the strength of its POS component \citep{lai2001,sandstrom2002,crutcher2004,girart2006,falceta-goncalves2008}.
In fact, this method is based on the {\angstructname}, which is obtained as the average of the {\dpsiname} over the positions.
The {\angstructname} is also used to study the magnetic field direction that can be inferred from different types of polarimetric measurements.
For example, \cite{Mao2010} computed the {\angstructname} in order to study the structures traced by the synchrotron Faraday rotation measures. 

\cite{serkowski1958} showed that the structure function of the Stokes parameters $Q$ and $U$ reaches a limit. 
When the area, considered to calculate the structure function, becomes too large and includes non-connected regions, the parameters become spatially decorrelated.
\cite{poidevin2010} reported about a similar behavior of the {\angstructname}.
The randomness of angles can be due not only to the physical decorrelation in the underlying pattern, but also to the noise of the measurement.
According to \cite{Hildebrand2009}, the {\angstructname} contains contributions of the large-scale and turbulent magnetic field components.
They have developed a method to estimate the strength of these components using the {\angstructname}.
The method has successfully been applied to polarimetry and interferometry data to characterize the magnetic turbulence power spectrum and magnetic field strength in molecular clouds \citep{houde2011b,houde2011a,houde2016}.
The authors claimed that its uncertainty can simply be calculated through the uncertainties of the angles used in the determination of the {\angstructname}. 

We have shown in Papers I and II that in order to accurately estimate the polarization fraction and polarization angle, one should take into account the full noise covariance matrix if possible.
In this work, we study the behavior of the bias on the {\dpsiname} knowing the full noise covariance matrix and the distribution of the true polarization angles.
We introduce new estimators of the {\dpsiname} and describe a method to evaluate an upper limit for the bias of the conventional estimator.

In Section \ref{sec:conventional} we introduce the notations and give the definition of the conventional estimator of the {\dpsiname} in terms of the Stokes parameters. 
In Section \ref{sec:bias} we demonstrate the peculiarity of the bias.
We also discuss the impact on the bias of the noise covariance matrix and of the distribution of the true polarization angles in the vicinity of the point of interest.
We address the reliability of the conventional uncertainty on {\dpsiname} as well.
In Section \ref{sec:recipes} we introduce alternative estimators and propose a method to evaluate the maximum bias of the conventional estimator for a given set of data.

\section{Conventional estimator of the {\dpsiname}}
\label{sec:conventional}
\subsection{Definition and notations}
\label{sec:def_and_notations}

A plane of the sky component of polarized radiation is characterized by the true, i.e. not affected by the measurement noise, polarization fraction
\begin{equation}
p_0 = \dfrac{\sqrt{Q_0^2+U_0^2}}{I_0} \, ,
\end{equation}
\label{eq:p}
and polarization orientation angle 
\begin{equation}
\angz = \dfrac{1}{2}\arctan(U_0,\, Q_0) \, ,
\label{eq:psi}
\end{equation}
where $I_0, \, Q_0, \, U_0$ are the true Stokes parameters that describe the intensity and the linear polarization of the incoming radiation. 
Function $\mathrm{arctan}$ takes two arguments in order to choose the correct quadrant when calculating the arctangent of the ratio $U/Q$. 

The true {\dpsiname} at the position $\mathbf{x}$, where $\mathbf{x}$ is the 2D coordinate in the POS, is defined as the root mean square over the $N(l)$ pairs of angles located within an area of radius $l$ around $\mathbf{x}$ (see Figure \ref{fig:delta_psi_scheme} for illustration):
\begin{equation}
\dpsiz (\mathbf{x},l) =  \sqrt{ \frac{1}{N(l)}  \sum_{i=1}^{N(l)} \left[ \angz(\mathbf{x}) - \angz (\mathbf{x}+\mathbf{l}_i ) \right]^2    }  \, .
\label{eq:dpsiz}
\end{equation}
$\dpsiz$ takes values between $0$ and $\pi/2$.
Note that it is also possible to consider only the angles contained in an annulus of a certain radius and width.
In that case $\dpsi = \dpsi(\mathbf{x}, l, \delta)$, where $\delta$ is the width of the annulus and $l$ is the lag. \\
When using the measured quantities, we will call this estimator the "conventional estimator" and denote it by $\cl$:
\begin{equation}
\cl (\mathbf{x},l) =  \sqrt{ \frac{1}{N(l)}  \sum_{i=1}^{N(l)} \left[ \ang(\mathbf{x}) - \ang (\mathbf{x}+\mathbf{l}_i ) \right]^2    } \, .
\label{eq:dpsi}
\end{equation}
The above formula takes the following form in terms of the Stokes $Q$ and $U$ parameters:
\begin{eqnarray}
\dpsi (\mathbf{x},l) = \Big[ \frac{1}{N(l)}\sum_{i=1}^{N(l)} \Big(\dfrac{1}{2} \arctan \lbrack U(\mx)Q(\mxl)-Q(\mx)U(\mxl),  \nonumber\\
Q(\mx)Q(\mxl)+U(\mx)U(\mxl) \rbrack \Big)^{2} \Big]^{1/2} \, .
\label{eq:dpsi_QU}
\end{eqnarray}
This equation is applicable to both $\cl$ and $\dpsiz$. 

Noise on any polarimetric measurement is characterized by a noise covariance matrix $\Sigma$.
The noise covariance matrix of a linear polarization measurement has the following form:
\begin{equation}
\Sigma \equiv \left(\begin{array}{ccc}
\sigma^2_{\rm I} & \sigma_{\rm IQ} & \sigma_{\rm IU} \\
\sigma_{\rm IQ} & \sigma^2_{\rm Q} & \sigma_{\rm QU} \\
\sigma_{\rm IU} & \sigma_{\rm QU} & \sigma^2_{\rm U} \\ 
\end{array}\right) \, ,
\label{eq:sigma}
\end{equation}
where $\sigma^2_{\rm X}$ ($X={I,\, Q, \, U}$) characterizes the noise level in the $X$ parameter (i.e. variance), and $\sigma_{\rm XY}$ ($Y={I,\, Q, \, U}$) characterizes the correlation between noise on $X$ and $Y$ (i.e. covariance). \\
As we are interested only in the angle measurements, the intensity is assumed to be known exactly, so that the noise covariance matrix can be reduced to:
\begin{equation}
\Sigma_p =  \left(\begin{array}{cc}
\sigma^2_{\rm Q} & \sigma_{\rm QU} \\
 \sigma_{\rm QU} & \sigma^2_{\rm U} \\ 
\end{array}\right)  \, .
\label{eq:sigma_simple_zero}
\end{equation}
It is possible to fully characterize $\Sigma_p$ using only two parameters \citep{Montier1}:
\begin{equation}
\varepsilon_{\rm eff}^2 = \frac{1 + \varepsilon^2
 + \sqrt{(\varepsilon^2-1)^2 + 4\rho^2\varepsilon^2}}
 {1 + \varepsilon^2 - \sqrt{(\varepsilon^2-1)^2 + 4\rho^2\varepsilon^2}}
\label{eq:epsi_eff}
\end{equation} 
and 
\begin{equation}
\label{eq:phi}
 \theta = \frac{1}{2} \mathrm{arctan}
 \left( \frac{2 \rho \varepsilon}{\varepsilon^2-1} \right) \, .
\end{equation}
Here $\eps$ and $\rho$ are the ellipticity and correlation between noises on $Q$ and $U$:
\begin{equation}
\rho = \frac{\sigma_{QU}}{\sigma_{Q}\sigma_{U}} \hspace{0.7cm} \rm{and} \hspace{0.7cm} \eps = \frac{\sigma_{U}}{\sigma_{Q}} \, .
\label{eq:epsrho}
\end{equation}
The reduced noise covariance matrix then takes the following form:
\begin{equation}
\Sigma_p =  \frac{\sigma^2_p}{\sqrt{1-\rho^2}}  \left(\begin{array}{cc}
1/ \varepsilon & \rho  \\
 \rho & \varepsilon \\ 
\end{array}\right) \, ,
\label{eq:sigma_simple}
\end{equation}
where $\sigma_p$ is a global polarization noise scaling factor, such that  $\det(\Sigma_p)=\sigma_p^4$ \citep{Montier1}.

The effective ellipticity $\epseff$ and the angle $\theta$ give the shape of the noise distribution in linear polarization, independently of the reference frame to which $Q$ and $U$ are attached.
 
In order to characterize the form of the noise covariance matrix, $3$ regimes of $\epseff$ are
considered in this study:
\begin{itemize}
\item the \textit{canonical} case:  $\epseff = 1$. This corresponds to the equality and independence between noise levels on $Q$ and $U$: $\qq=\uu$, $\sigma_{\rm UQ}=\sigma_{\rm UQ}=0$;  
\item the \textit{low} regime: $1\,{\le}\,\varepsilon_{\rm eff}\,{<}\,1.1$. This means that the differences and/or correlations between noise levels on $Q$ and $U$ are small;
\item the {\it extreme} regime: $1.1\,{\le}\,\varepsilon_{\rm eff}\,{<}\,2$. This means that the differences and/or correlations between noise levels on $Q$ and $U$ are large.
\end{itemize}

\subsection{Monte Carlo simulations}
\label{sec:mc_sim}

\begin{figure}[!h!t]
\begin{center}
\includegraphics[width=9cm]{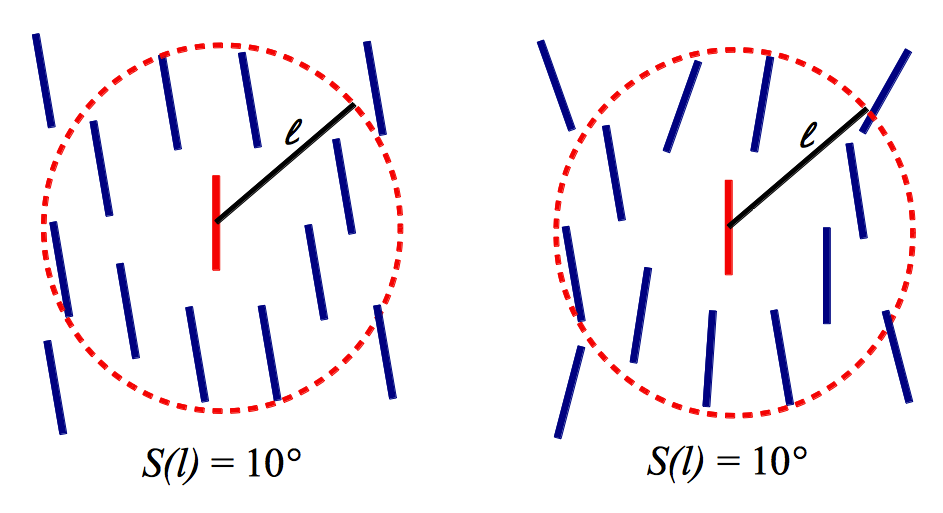} 
\caption{A schematic view of the simulated configuration of polarization orientations. The {\dpsiname} is calculated at the position of the red line segment within the red-dotted circle of radius $\textit{l}$. Left: "uniform" configuration. Right: "random" configuration. Both cases give $\dpsi=10^{\circ}$.}
\label{fig:delta_psi_scheme}
\end{center}
\end{figure}

In order to characterize the bias on the {\dpsiname}, we perform Monte Carlo (MC) simulations.
We build numerical distribution functions (DFs) of $\cl$ using the following set of basic assumptions:
\begin{enumerate}
\item We consider $10$ pixels: $1$ central pixel and $9$ adjacent pixels to be contained within a circle of radius $l$, as shown in Figure \ref{fig:delta_psi_scheme}. In a regularly-gridded map there are $8$ adjacent pixels, but a small difference (by $1$ or $2$) in the number of pixels does not affect the results of our simulations.  
\item All pixels have the same true polarization fraction $p_0=0.1$ and the same noise covariance matrix $\Sigma_p$. The latter assumption seems to be reasonable because $\dpsi$ is usually calculated inside small areas, where the instrumental noise does not change much.
\item We perform $N_{\rm MC}=10^6$ noise realizations at each run (i.e. for each simulated configuration, including the signal-to-noise ratio ({\sn}), the true value, the shape of the noise covariance matrix and the true polarization angles).
\item We consider Gaussian noise on $Q$ and $U$ with a noise covariance matrix $\Sigma_p$. 
\item We vary the {\sn} of $p$ between $0.1$ and $30$. We set $\sigma_p=p_0/(\sn)$ to be used in Equation \ref{eq:sigma_simple}. 
\item We vary $\rho$ in the range $[-0.5, \, 0.5]$ and $\epsilon$ in the range $[0.5,\, 2]$. The {\low} regime is obtained when using $\rho \simeq 0$ and $\eps \simeq 1$; other cases (with $\epsilon \leq 0.9$, and $\epsilon \geq 1.1$ and $\rho \geq \vert 0.05 \vert$) give the {\extreme} regime of $\epseff$.
\end{enumerate}

We use $\ang_{0,i}$ to denote the true polarization angle for pixel $i$, and consider two cases of the configuration: the "uniform" and the "random" configurations.
In the "uniform" configuration, all angles $\ang_{0,i}$ are the same for $i \ \in \,[1,9]$, while $\ang_{0,0}$ is calculated as:
\begin{equation}
\ang_{0,0} = \ang_{0,i} - \dpsiz \, . 
\end{equation}
In the "random" configuration $\ang_{0,i}$ for $i \ \in \ [1,9]$ are generated randomly and $\ang_{0,0}$ is selected from a series of random values  to obtain $\dpsiz$ with $(10^{-5})^{\circ}$ precision using Equation \ref{eq:dpsiz} at each run.
Examples of both configurations, "uniform" and "random", are illustrated on left and right panels in Figure \ref{fig:delta_psi_scheme}, respectively. 
There are $10$ representative sets of the true angles for each configuration and the true {\dpsiname}.
They are obtained by varying $\ang_{0,0}$ from $0$ to $\pi/2$ with $10^{\circ}$ ($\pi/18$) step for the "uniform" configuration and by generating additional sets for the "random" configuration. 

\begin{figure}[!h!t]
\begin{center}
\includegraphics[width=9cm]{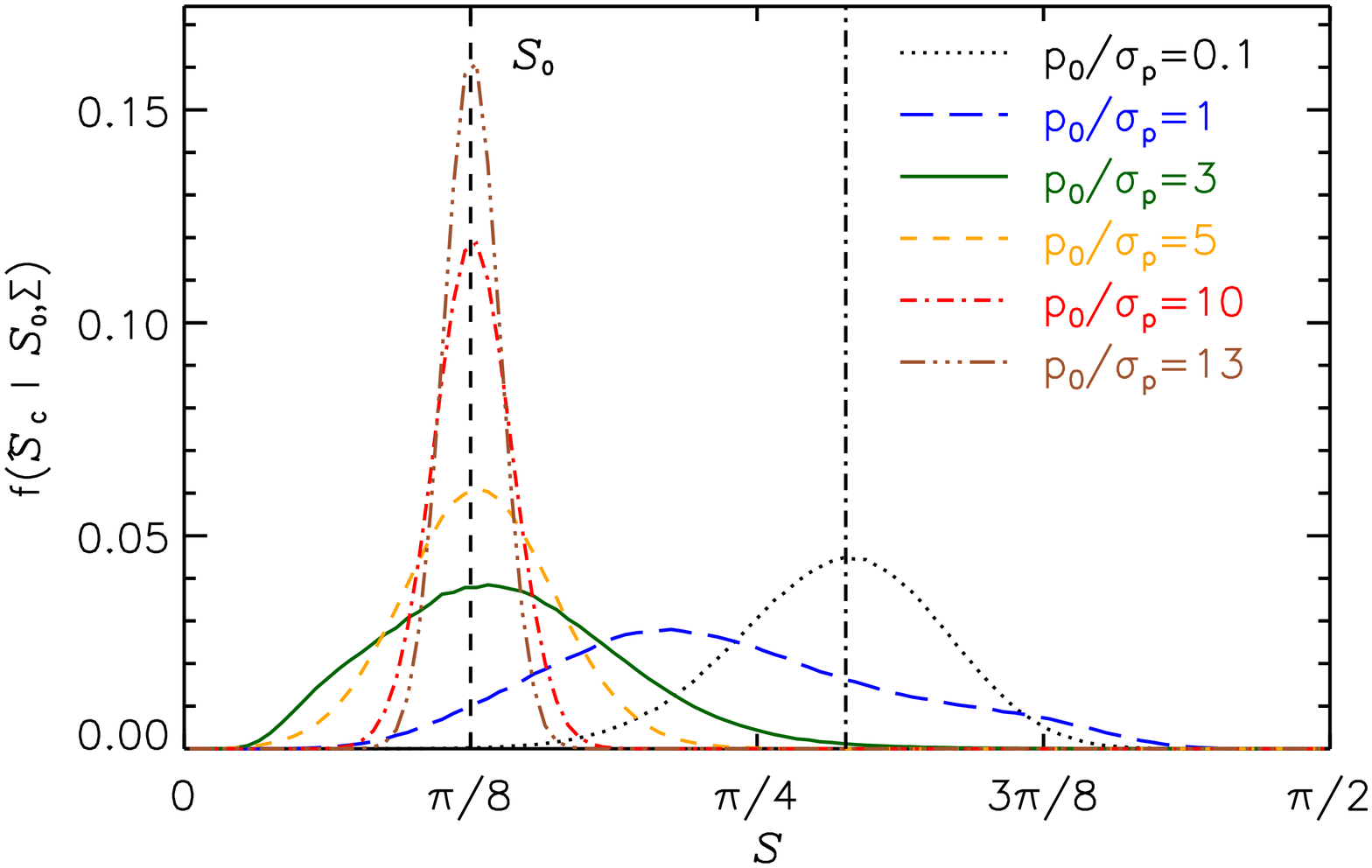}
\includegraphics[width=9cm]{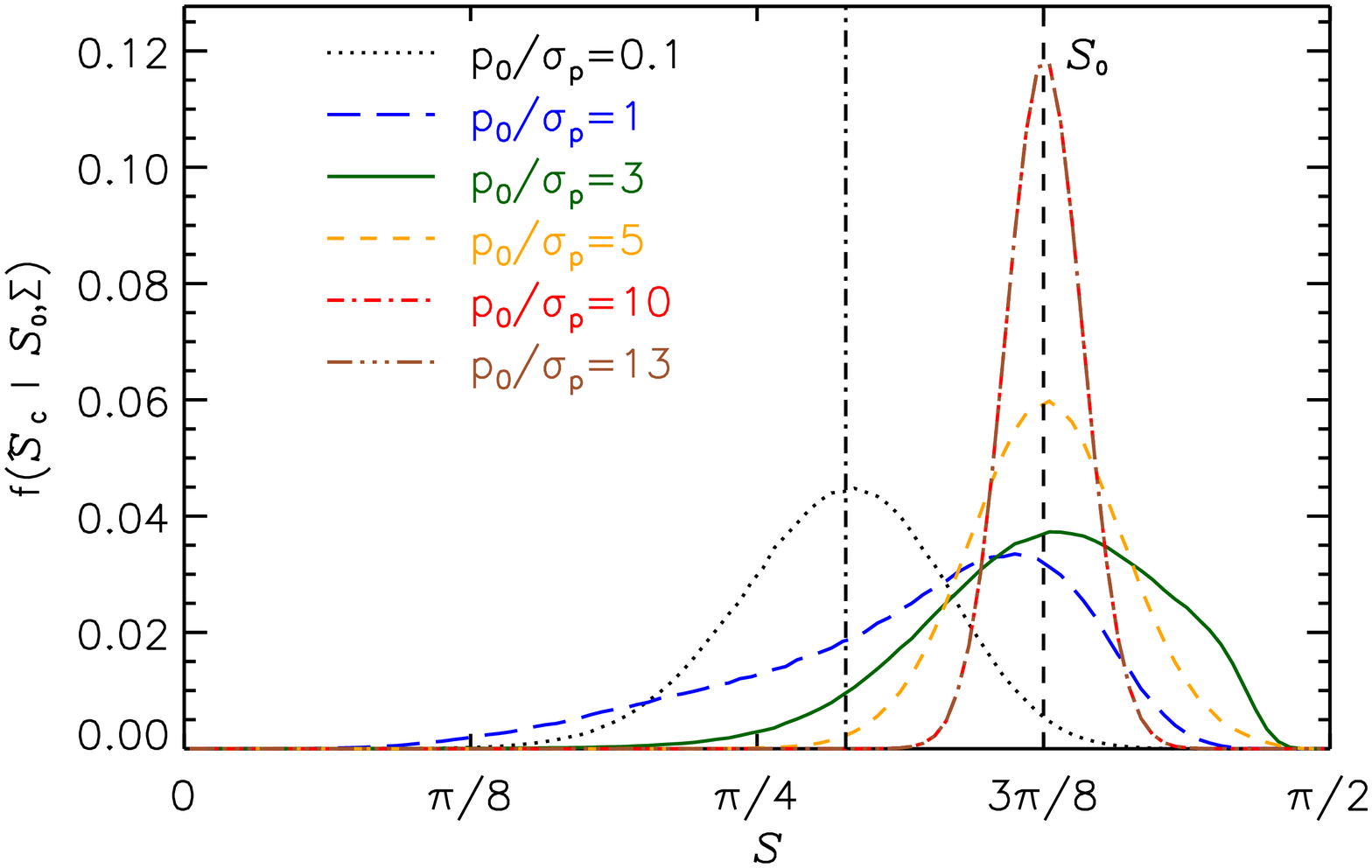}
\caption{Examples of the simulated distribution functions of the conventional estimator of the dispersion function $\cl$ for different {\sn}s of $p$ in the {\canonical} case of the noise covariance matrix. Top: $\dpsiz = \pi/8$, bottom: $\dpsiz = 3\pi/8$. The vertical dashed line shows the true value, and the vertical dash-dotted line shows the value of $\pi/\sqrt{12}$.}
\label{fig:examples}
\end{center}
\end{figure}

Once $\ang_{0,0}$ and $\ang_{0,i}$ are obtained, the following transformation is performed in order to get the corresponding $Q$ and $U$ parameters:
\begin{eqnarray}
Q_{0,i} &=& p_0 \, I_0 \, \cos(2\ang_{0,i})  \, , \ i \in[0,9] \, , \\
U_{0,i} & =& p_0 \, I_0 \, \sin(2\ang_{0,i}) \, , \ i \in[0,9] \, ,
\label{eq:ppsi2qu}
\end{eqnarray}
with $I_0 = 1$.
Random Gaussian noise is generated for each pixel for $Q$ and $U$ according to the noise covariance matrix and is added to the true values to obtain the simulated Stokes parameters for each pixel.
The simulated measured {\dpsiname} $\cl$ is calculated using Equation \ref{eq:dpsi_QU}. \\
Once we have the simulated sample of $10^6$ values of $\cl$ for the given $\dpsiz$, the configuration of the true angles and the noise level, we can build numerical DFs, which we denote as $f(\cl \, | \, \dpsiz, \Sigma)$.
The shape of the  DF for the given noise levels in the {\canonical} case of the noise covariance matrix and in the "uniform" configuration of the true angles is illustrated in Figure \ref{fig:examples}. 
At very low {\sn}s, the distribution function peaks at $\pi/\sqrt{12}$, regardless of $\dpsiz$.
The value $\randval$ ($\simeq 51.96^{\circ}$) corresponds to the result of $\dpsi$ with purely random distribution of angles.
In fact, for a pair of angles in the range $[-\pi/2,\, \pi/2]$, their absolute difference is distributed uniformly in the range $[0, \, \pi/2]$.
The root mean square of this distribution gives $\randval$.

\section{Bias analysis}
\label{sec:bias}

In the following, the bias on $\dpsi$ is calculated as follows:
\begin{equation}
Bias = \frac{1}{N_{MC}}\sum_{k=1}^{N_{MC}} \hat{\dpsi}_{C,k} - \dpsiz = \langle \cl \rangle - \dpsiz \, ,
\end{equation}
where $\hat{\dpsi}_{C,k}$ is a realization of the conventional estimator of $\dpsi$.
We study different origins of the bias on $\cl$ by comparing the contributions of the biases due to the following parameters that affect its estimation: the true value $\dpsiz$ ($Bias_{\dpsiz}$), the shape of the noise covariance matrix ($Bias_{\dpsiz, \Sigma}$), the distribution of the true angles ($Bias_{\dpsiz, \angz}$) and the joint impact of these parameters ($\biasjoint$).

\subsection{Impact of the true value $\dpsiz$}
\label{sec:bias_dpsiz}
\begin{figure}[!h!t]
\begin{center}
\includegraphics[width=9cm]{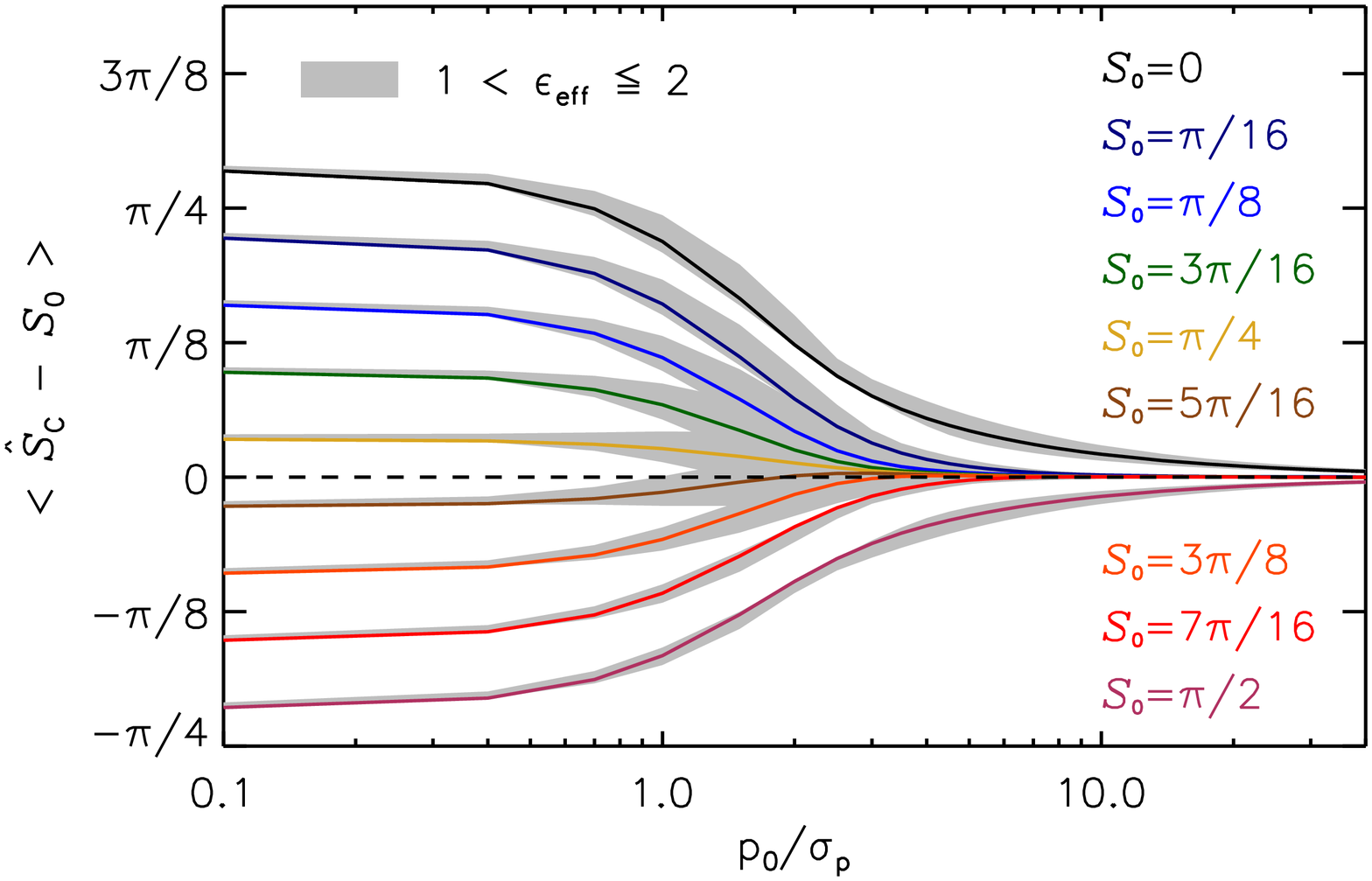}    
\caption{The average bias on $10^{6}$ MC noise realizations for the conventional estimator $\cl$ for different true values $\dpsiz$ as a function of {\sn}: in the {\canonical} case of the noise covariance matrix configuration ($\epseff=1$) - colored plain curves and in the \textit{extreme} regime ($\epseff$ up to $2$). The colored curves are shown from top to bottom in the same order as the legend lines on the right part of the Figure. The {\low} regime regions are invisible at the current plot scale and coincides with colored curves. The dashed line represents the "zero bias" level.}
\label{fig:bias_example}
\end{center}
\end{figure}

We calculate the average statistical bias induced by noise and the true value, $\biascl$, in the case with $\epseff=1$ and "uniform" configuration of the true angles.
Figure \ref{fig:bias_example} represents $\biascl$ (in colored plain curves) as a function of {\sn}, for values of $\dpsiz$ ranging from $0$ to $\pi/2$ in steps of $\pi/16$ ($11.25^{\circ}$).
If the {\sn} is high, $\cl$ corresponds to $\dpsiz$, whereas if {\sn} is low, $\cl$ does not represent $\dpsiz$.
The closer $\dpsiz$ to the bounds ($0$ or $\pi/2$), the larger the bias $Bias_{\dpsiz}$, even at high {\sn} ($\snrpo>10$).
The largest bias occurs in the case where $\dpsiz=0$, which is the most remote value from $\randval$ (where $\randval$ is the result for $\cl$ if the orientation angles are random).
Also, the conventional estimator $\cl$ can be ambiguous if it gives results close to $\pi/\sqrt{12}$. \\
In the presence of noise, $\cl$ is biased, though not necessarily positively biased, whereas the polarization fraction $p$ is always positively biased \citep{Montier1}.
For a true value of  $\dpsiz$ lower than $\randval$, the measured $\cl$ is positively biased, while it has negative bias for $\dpsiz$ larger than $\randval$.

\subsection{Impact of the (Q,U) effective ellipticity}
\label{sec:cov_matrix}

 \cite{Montier1} showed that the shape of the noise covariance matrix associated with a polarization measurement affects the bias on the polarization fraction $p$ and angle $\ang$.
Here we study the impact of the shape of the noise covariance matrix on the bias of the conventional estimator of the {\dpsiname} and evaluate under what conditions the assumption of non-correlated noise (i.e. $\epseff=1$) can be justified.
For this purpose, we run the MC simulations as described in Section \ref{sec:mc_sim} in the three cases of the effective ellipticity and in the "uniform" configuration of the true angles.

We show in Figure \ref{fig:bias_example} the statistical bias of $\cl$ depending both on the true value and on the shape of the noise covariance matrix, $\biassig$, as a function of {\sn} and for different true values $\dpsiz$.
In the \textit{low} regime the shape of $\Sigma_p$ has practically no effect on the bias: the corresponding dispersion can not be seen in the Figure as it coincides with the {\canonical} case curves.
A dispersion in the initial bias $\biascl$ (corresonding to the amplitude of the gray areas) appears if there are important asymmetries in the shape of $\Sigma_p$, i.e. in the \textit{extreme} regime.
Note that these asymmetries may either increase or decrease the statistical bias: $<\cl - \dpsiz> $ in the gray areas are higher or lower than the colored curves, i. e. closer to or farther from the "zero bias" line, that occurs for $\randval$ in the {\canonical} case and shown by the dashed line in the Figure.
If the true {\dpsiname} is close to $\randval$, i.e. close to the "zero bias" line, $Bias_{\dpsiz, \, \Sigma}$ is significant with respect to $Bias_{\dpsiz}$ (for $\dpsiz = 3\pi/16, \, \pi/4, \, 5\pi/16, \, 3\pi/8$). 
If $\dpsiz$ is very different from $\randval$, i.e. remote from the "zero bias" line, both $\biascl$ and $\biassig$ become comparable for $S/N \geq 3$ (for $\dpsiz = 0, \, \pi/16,\, \pi/8,\, 7\pi/16, \, \pi/2$).

\begin{figure}[!h!t]
\includegraphics[width=9cm]{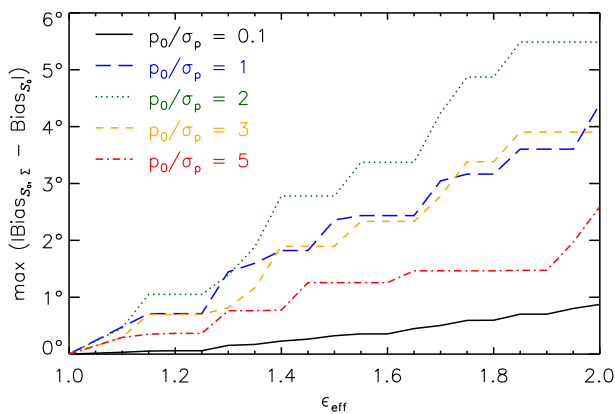}
\caption{The maximum absolute deviation of the bias induced by variations of the effective ellipticity between noise in ($Q$, $U$) and the true value $\dpsiz$, $Bias_{\dpsiz,\, \Sigma}$, from the bias induced by only the true value in the {\canonical} case, $Bias_{\dpsiz}$ as a function of the effective ellipticity for different {\sn}.}
\label{fig:cov_matrix_bias_deviation}
\end{figure}

The dispersion in the bias $\biassig$ reaches its maximum at intermediate {\sn} ($\snrpo \ \in \ [1, 3]$).
At low {\sn} ($\snrpo < 0.5$), there is almost no impact of the shape of the noise covariance matrix on the bias and we observe only the bias due to $\dpsiz$: the dispersion of $\biassig$ is much smaller than the level of $\biascl$.
When the noise level is too high, it dominates any other effect.
At high {\sn}, the noise level is low, so the estimation becomes accurate enough to become independent of the shape of the noise covariance matrix.
Figure \ref{fig:cov_matrix_bias_deviation} shows the maximum absolute deviation of $\biassig$ from  $\biascl$ over all possible values of $\dpsiz$ as a function of $\epseff$. 
The maximum deviation increases progressively with $\epseff$ and is the largest at $\snrpo=2$ with $\max(\vert < Bias_{\dpsiz, \, \Sigma} - \biascl > \vert) = 5.3^{\circ}$ ($\pi/34$).

Thus, the shape of the noise covariance matrix can significantly impact the bias on the {\dpsiname}.
In the \textit{extreme} regime and intermediate {\sn}, for the true values close to $\randval$, the bias induced by the ellipticity and/or correlation between noise levels on $Q$ and $U$ is of the same order as the bias due to $\dpsiz$ in the {\canonical} case (as for the values of $\dpsiz$ between $3\pi/16$ to $3\pi/8$ in Figure \ref{fig:bias_example}): the width of the gray areas is comparable to the amplitude of the colored curves. 
Nevertheless, in the case where irregularities of the noise covariance matrix depart by less than $10\%$ from the {\canonical} case, i.e. in the {\low} regime, the impact of the asymmetry in the shape of the noise covariance matrix on the bias of $\cl$ is negligible (the amplitude of the deviation from the bias in the {\canonical} case $Bias_{\dpsiz}$ is very low and is not represented in the Figure).

\subsection{Impact of the true angles distribution}
\label{sec:true_angle_eps1}

\begin{figure}[!h!t]
\includegraphics[width=9cm]{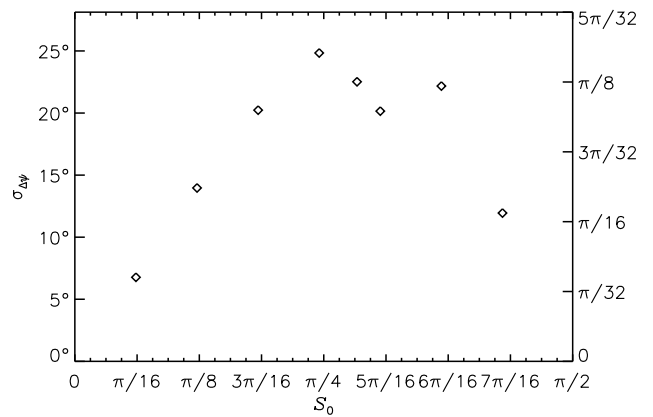}
\caption{The standard deviation of the difference between the true angle $\psi_{0,0}$ and the true angles $\psi_{0,i} , \, i \, \in \, [1,9]$ as a function of the true {\dpsiname} $\dpsiz$ in the {\canonical} case of the noise covariance matrix and "random" configuration of the true angles.}
\label{fig:dpsi_disp}
\end{figure}

A multitude of different combinations of the true polarization angles $\ang_{0,i}$ can yield the same value $\dpsiz$.
We study to which extent the {\dpsiname} can be affected by the configuration of the true angles.
We compare the bias induced by the different configurations of the angles $Bias_{\dpsiz,\angz}$ to the bias due to the true value $Bias_{\dpsiz}$ in the "uniform" configuration (seen in Section \ref{sec:bias_dpsiz}).
For this purpose, we perform simulations in the {\canonical} case of the noise covariance matrix for the $10$ simulated combinations of the true polarization angles in each of the configurations ("random" and "uniform"). 
Figure \ref{fig:dpsi_disp} shows the dispersion $\sigma_{\Delta \psi}$ of the differences between angles of the central pixel and of the neighbor pixels $\Delta \psi_{0,\,i}$ for $i \, \in \, [1,9]$ as a function of $\dpsiz$ in the {\canonical} case of the noise covariance matrix and the "random" configuration of the true angles.
The dispersion of the angles that give the value $\dpsiz=\randval$ is also shown (the point between $\dpsiz=\pi/4$ and $\dpsiz = 5\pi/16$).
Note that by construction, random distributions of the true angles that give $\dpsiz = 0$ and $\dpsiz=\pi/2$ do not exist. 
Also, the closer $\dpsiz$ to these values ($0$ and $\pi/2)$, the smaller the dispersion because there are less possible combinations of $\Delta \psi_{0,\, i}$.

\begin{figure}[!h!t]
\includegraphics[width=9cm]{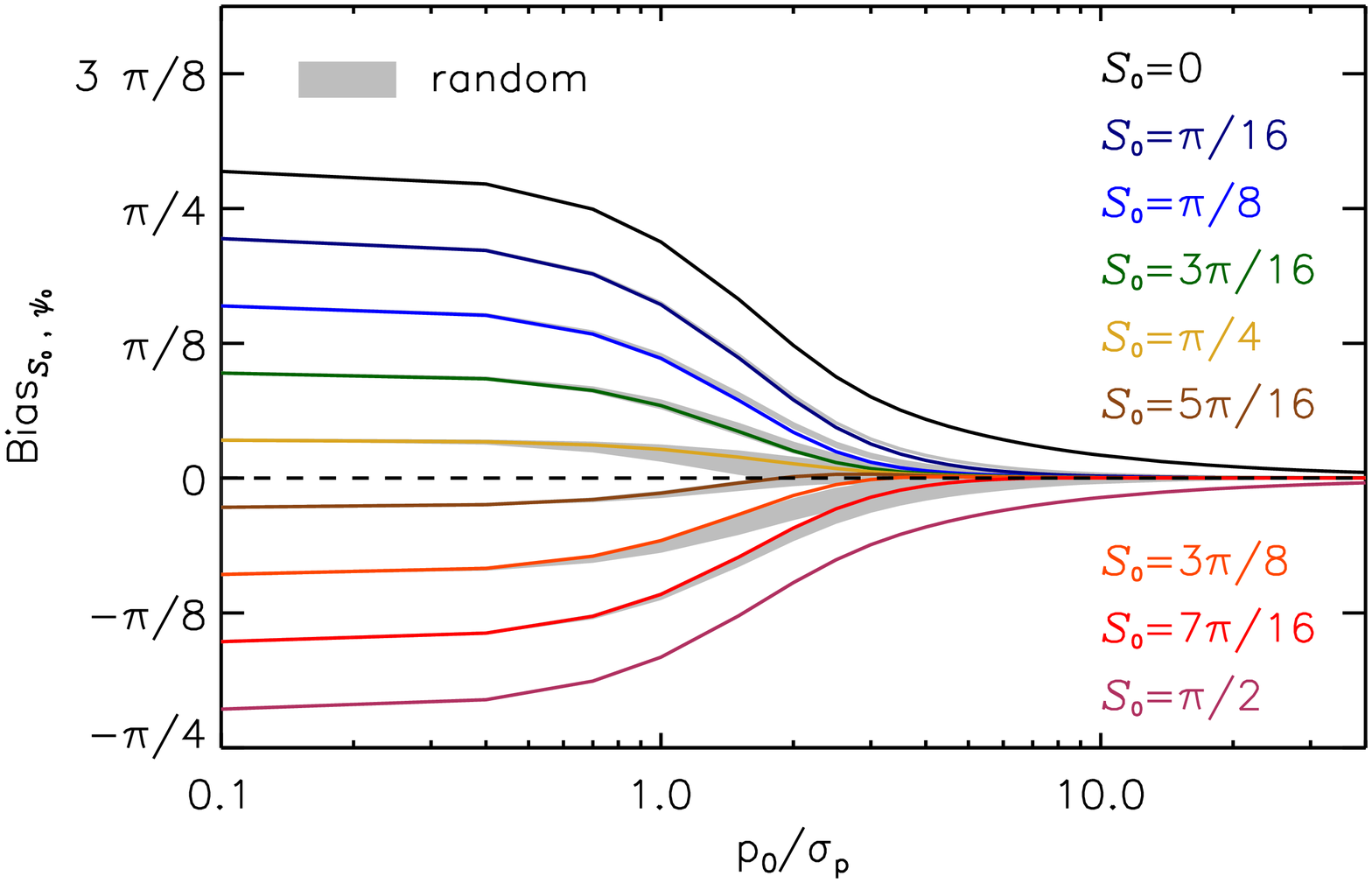}
\caption{The average bias on $10^6$ MC noise simulations on $\cl$ in the "uniform" distribution of the true angles $\psi(\mxl)$ (colored curves) and the dispersion of the average bias in the "random" distribution of the true angles (gray areas) in the {\canonical} case of the noise covariance matrix ($\epseff = 1$). The colored curves are shown from top to bottom in the same order as the legend lines on the right part of the Figure. The dashed line represents the "zero bias" level.}
\label{fig:eps1_angs_distribution}
\end{figure}

In Figure \ref{fig:eps1_angs_distribution} we show the examples of the statistical bias $\biasang$ obtained in both configurations of the true angles. 
The different realizations of the "uniform" configuration in the {\canonical} regime does not bring any contribution to the bias $\biascl$ obtained in the {\canonical} case of the noise covariance matrix and fully reproduce the colored curves of Figure \ref{fig:bias_example}. 
But when the distribution of the angles deviates from  uniformity and becomes random, variations in the bias appear. 
In fact, each pair of angles $(\psi_{0,0}, \psi_{0,i})$ has its proper $\Delta \ang_{0,i} = \psi_{0,0} - \psi_{0,i}$ and  only their mean squared sum gives $\dpsiz$. 
In the presence of noise, $\Delta \ang_{0,i}^2$ becomes  biased.
The sum of the biased quantities results in the dispersion of the total bias on $\cl$.

Similarly to the case of the bias induced by both the true value and the shape of the noise covariance matrix $Bias_{\dpsiz, \, \Sigma}$, the dispersion in the bias due to the true value and the true angles distribution $\biasang$ increases at intermediate {\sn} and diminishes at low and high {\sn}, for the same reason discussed in Section \ref{sec:cov_matrix} (gray areas become larger at intermediate {\sn} in Figure \ref{fig:eps1_angs_distribution}).
$\dpsiz = \pi/4$ opens the widest range of possible $\Delta \ang_{0,i}$, ensuring the largest dispersion of values (Figure \ref{fig:dpsi_disp}).
Thankfully, this value has a small bias due to $\dpsiz$: the corresponding colored curve in Figure \ref{fig:eps1_angs_distribution} is close to the "zero bias" level even at low {\sn}.
At $p_0/\sigp = 2$, the maximum dispersion of the bias for $\dpsiz = \pi/4$ is almost $4^{\circ}$ ($\simeq \pi/45$, corresponding to the width of the grey area) when the angles are distributed randomly, whereas the bias due only to noise is $0.8^{\circ}$ ($\pi/225$).

In the {\canonical} regime, the impact of the distribution of the angles used to calculate $\cl$ can be of the order of few degrees in the worst case, i.e. if the true angles are distributed quasi-randomly. 
However, in real observational data one would expect the polarization angles to be distributed neither uniformly nor randomly but within a particular structure in-between these two extreme configurations.
The bias will increase with the number of pairs of angles ($\psi(\mx), \psi(\mxl)$) used for the computation of $\dpsi$, i.e. with the radius $l$, as reported by \cite{serkowski1958}.
The {\angstructname} of $Q$ and $U$ obtained by \cite{serkowski1958} in the Perseus Double Cluster reached a limit when taking a radius larger than $12.8'$ with 24 pairs of parameters taken into account.\\
In the {\canonical} case of the noise covariance matrix, the impact of the true angles on the bias on $\cl$ can be neglected if a reasonable radius (or lag and width) with respect to the resolution of the data, is considered in the calculation.
E.g., \cite{planck2014-xix} calculated the {\dpsiname} at a lag of $30'$ with $30'$ width which corresponds to $28$ orientation angles at $1^{\circ}$ degree resolution. 

\subsection{Joint impact of the $(Q,U)$ ellipticity and of the distribution of the true angles}
\label{sec: true_angle_alleps}

In this section, we study the simultaneous impact of the shape of the noise covariance matrix and of the distribution of the true angles on the estimation of $\cl$. \\
\cite{Montier1} showed that if the effective ellipticity between noise levels on $Q$ and $U$ differs from $1$, then the bias on the polarization angle $\ang$ oscillates depending on the true angle $\ang_0$. 
The period of the oscillations is about $\pi/2$ (see their Figure $14$).
Thus, if there is a true difference $\Delta \psi_{0,i} = \pi/4$ between angles $\ang_0(\mathbf{x})$ and $\ang_0(\mathbf{x}+\mathbf{l_i})$, their respective biases can maximize the total difference $\Delta \psi_i$ for some pairs.
Note if the noise components on $Q$ and $U$ are correlated (i.e. $\rho \neq 0$), $\dpsiz= \pi/4$ will remain the value that yields the largest relative bias, while only the overall pattern would be shifted along $\psi_0$. 

\begin{figure}[h!]
  \begin{minipage}[c]{0.58\textwidth}
    \includegraphics[width=9cm]{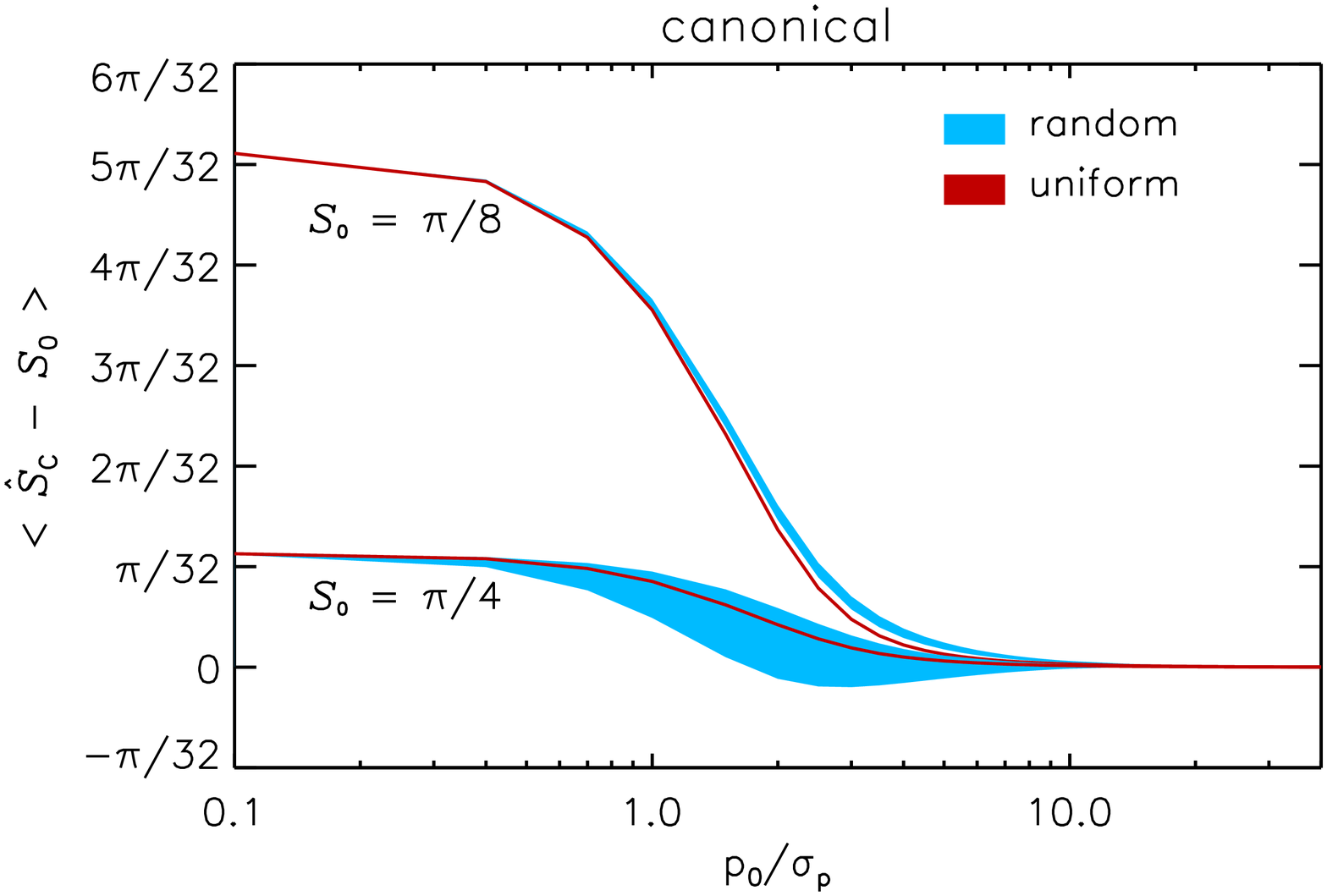}
        \includegraphics[width=9cm]{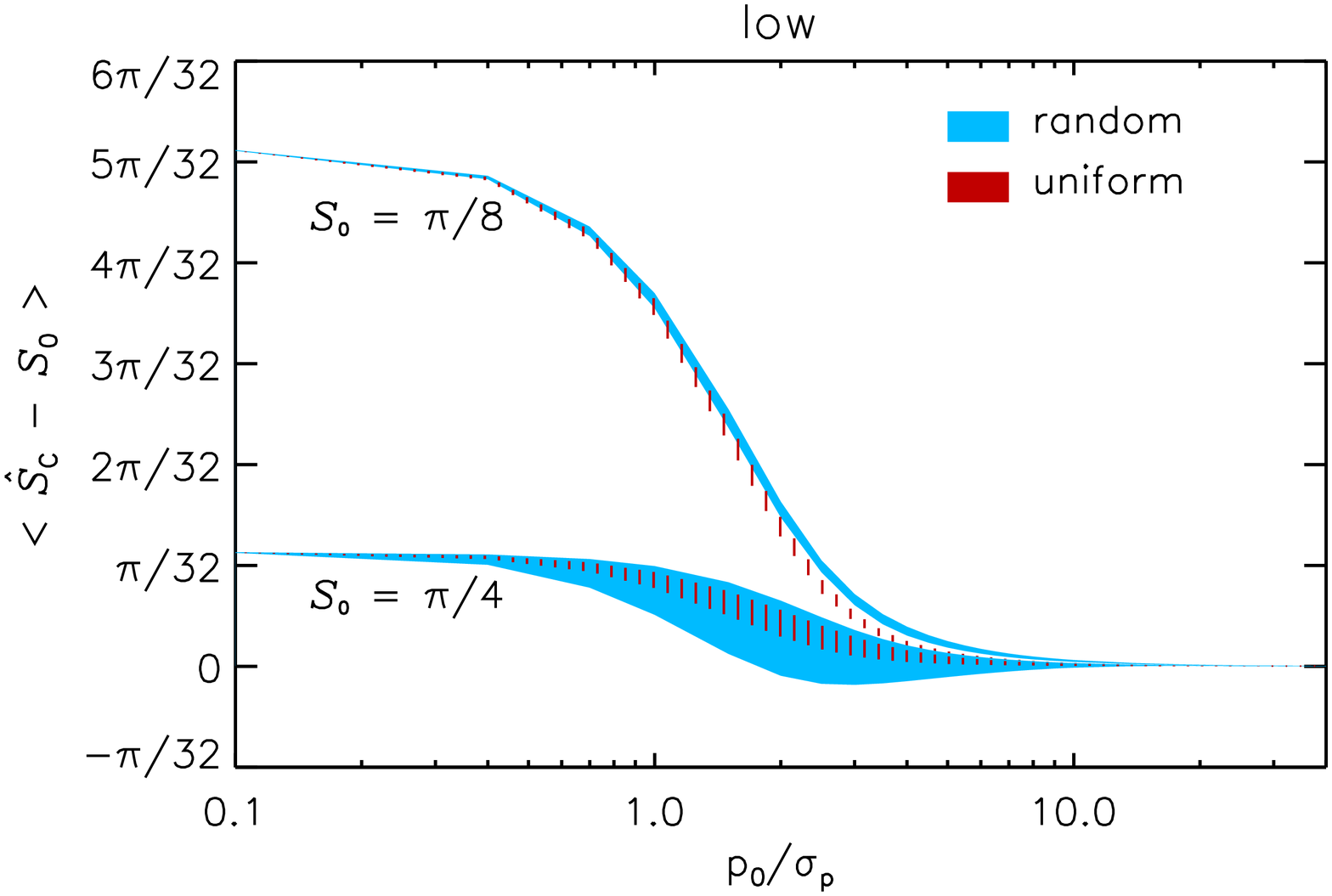}
    \includegraphics[width=9cm]{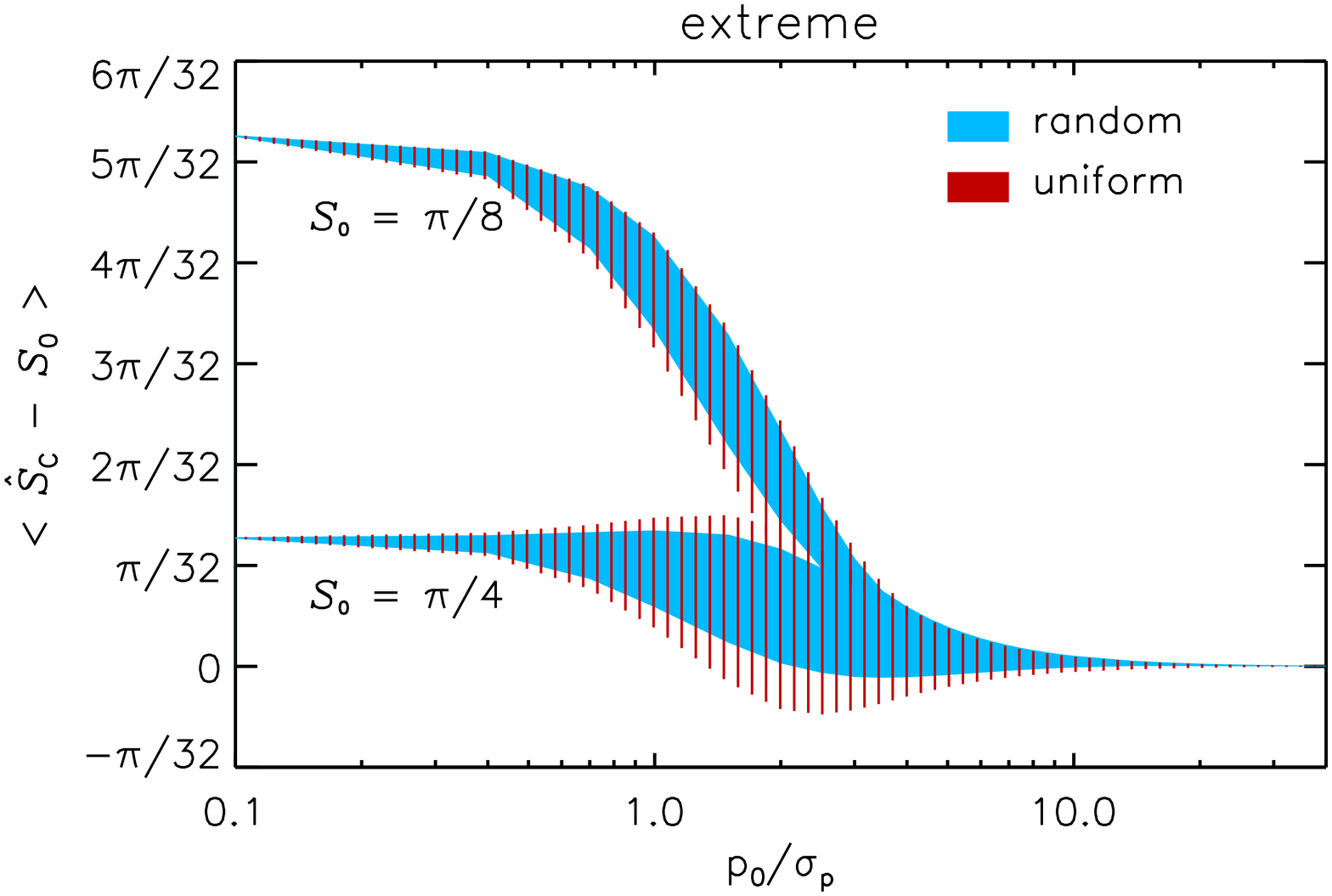}
  \end{minipage}\hfill
  \begin{minipage}[c]{9cm}
    \caption{
             The average bias on $10^6$ MC realizations of the conventional estimator of the {\dpsiname}. Blue filled and red hashed areas delimit dispersion over $10$ different sets of the true angles distributed randomly (blue) and uniformly (red) in three regimes of the shape of the noise covariance matrix, from top to bottom: {\canonical}, {\low}, {\extreme} regimes.
    } 
    \label{fig:angs_and_covmatrix}
  \end{minipage}
\end{figure}

We run numerical simulations for the true value $\dpsiz=\pi/4$ that would maximize the bias between pairs of angles in the case $\epseff \neq 1$.
We also explore $\dpsiz = \pi/8$ for illustration purposes.
We show in Figure \ref{fig:angs_and_covmatrix} the average bias for the "uniform" and "random" configuration of the true angles in the {\canonical}, {\low} and {\extreme} regimes. 
For $\epseff \neq 1$ (i.e. in the {\low} and {\extreme} regimes), the dispersion in the bias appears for both configurations, which is represented by the vertical width of the curves in the middle and bottom panels in Figure \ref{fig:angs_and_covmatrix}.
In the {\low} regime, the "uniform" configuration of the true angles gives a dispersion that is lower than the dispersion in the "random" configuration for $\dpsiz = \pi/4$.
However, in the {\extreme} regime the situation is the opposite. 
This can be due to the fact that in the "uniform" configuration, the imposed $\dpsiz$ is valid for every pair of angles, thus giving $\Delta \ang_{0,i}= \dpsiz$, so that the relative bias between angles in a pair is maximized for some of the combinations.
When angles are distributed randomly, $\dpsiz$ is ensured for the ensemble, but not for each pair: the pairs of angles with little relative bias diminish the final result. 
For $\dpsiz = \pi/8$ and for other $\dpsiz \neq \pi/4$ (not shown here), the observed difference between the "random" and "uniform" cases in the three regimes of $\epseff$ is less prominent than for $\dpsiz = \pi/4$, but the overall behavior does not change.

The joint impact of the distribution of the true angles and the shape of the noise covariance matrix on the bias of $\cl$ is high at intermediate {\sn}.  
In the {\extreme} regime and in the "uniform" configuration, the dispersion in the bias with respect to the {\canonical} case reaches its maximum of $10.1^{\circ}$ ($\simeq \pi/18$) at $p_0/\sigp=2$. 
This is not far from the value of the dispersion due to variations of the effective ellipticity only, given by the width of the grey area for $\dpsiz=\pi/4$ in Figure \ref{fig:bias_example} ($8.9^{\circ}$, $\simeq \pi/20$).
On the contrary, the dispersion in the bias in the "random" configuration gives only $6.4^{\circ}$ ($\simeq \pi/28$) in the same {\sn} range.
Thus, if the angles become random, it has little impact on the bias in the {\extreme} regime.
In the {\low} regime and "random" configuration, the maximum dispersion in the bias is $4.2^{\circ}$ ($\simeq \pi/43$) at $\snrpo = 2$, while it is equal to $1.5^{\circ}$ ($\pi/120$) in the "uniform" configuration.
Such a behavior of the bias can have a particularly strong impact on the estimation of $\dpsi$.
Consider a polarization pattern where angles become decorrelated with the distance: close to the pixel of interest, angles are more or less similar, becoming "random" with increasing distance from it.
The angles close to the pixel for which the {\dpsiname} is calculated, will be affected more by the bias (positive or negative) due to the distribution of true angles than those which are farther. 
This would lead to a non-homogeneity in the estimation of the {\dpsiname} in both {\low} and {\extreme} regimes of the noise covariance matrix. 
Such an issue will not arise if one considers the {\dpsiname} calculated at a given lag, $\dpsi(\mathbf{x}, l, \delta)$, and if the width of the annulus is small compared to the typical scale for decorrelation of angles

\subsection{Conventional uncertainties}
\label{sec:uncertainty}

\begin{figure}[h!]
\begin{center}
\includegraphics[width=8.5cm]{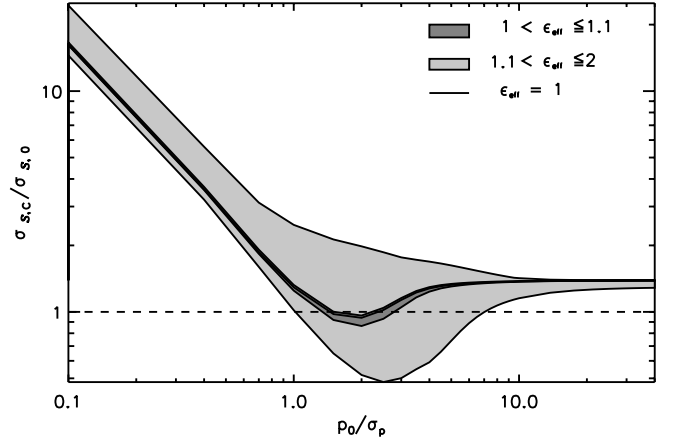}
\caption{The ratio between the conventional uncertainty and the true uncertainty of {\dpsiname} for different configurations of the noise covariance matrix. The dashed line represents the value of $1$.}
\label{fig:sig_ratio}
\end{center}
\end{figure}

As soon as the uncertainties of each of the angles $\psi(\mathbf{x})$ and $\psi(\mathbf{x+l_{i}})$ can be derived, one can obtain an estimate of the uncertainty on $\cl$ using the partial derivatives method.
Such an estimator of the uncertainty will be called the "conventional" estimator hereafter.
The conventional uncertainty of $\cl$ is given by (see Appendix \ref{app_derivatives} for derivation):
\begin{eqnarray}
\sigcl  =  \dfrac {1}{N \dpsi (\mathbf{x},l)}  && \Big[(\sum_{i=1}^{N}[ \psi (\mathbf{x}) -  \psi (\mathbf{x} +\mathbf{l}_{i} ) ])^{2} \sigma^{2}_{\psi(\mathbf{x})} \\
 &&+ \sum_{i=1}^{N} [\psi (\mathbf{x}) -  \psi (\mathbf{x} +\mathbf{l}_{i} ) ]^{2} \sigma^{2}_{\psi(\mathbf{x+l_{i}})} \Big]^{1/2} \, .
\label{eq:uncertainty}
\end{eqnarray}
Although the conventional method is limited to relatively high {\sn}s to ensure small deviations from the true value, it is the easiest method to derive an uncertainty on $\cl$ once the data and the associated noise information for each component are available. 
In order to quantify to which extent the conventional uncertainty can be reliable, we compare it to the uncertainty on $\cl$ given by the standard deviation of the distribution, denoted by $\sigma_{\dpsi,0}$.
The ratio of these uncertainties is shown in Figure \ref{fig:sig_ratio} in the {\canonical}, {\low}, and {\extreme} regimes.
Uncertainties on the angles, $\sigma^{2}_{\psi(\mathbf{x})}$, $\sigma^{2}_{\psi(\mathbf{x+l_{i}})}$, used in the determination of $\sigcl$, are also calculated by the conventional method \citep{Montier1} using $Q$ and $U$ and noise covariance matrices $\Sigma_{p,i}$ of each pixel. 
Then, one should note that $\sigma_{\psi(\mathbf{x})}$ and $\sigma_{\psi(\mathbf{x+l_{i}})}$ are themselves subject to the limitation of the derivatives method.

At low {\sn} ($\snrpo<1$), the estimate of the uncertainty using the conventional method is very inaccurate.
In the {\canonical} case of the noise covariance matrix, $\sigcl$ rapidly converges toward the true uncertainty and becomes compatible within $10\,\%$ in the range $\snrpo \ \in \ [1,\,3]$.
Then it increases at higher $S/N$ and overestimates the uncertainty on {\dpsiname} up to $38\, \%$ at high (larger than $20$) $S/N$ of $p$. 
The ratio does not converge to $1$ at high {\sn}s.
In the case of more complex shapes of the noise covariance matrix, $\sigcl$ can deviate from the true value by a factor of $2$ at {\sn}s ranging between $1$ and $10$.
At {\sn} larger than $10$, the ellipticity and correlation between $Q$ and $U$ do not affect the estimation of the uncertainty and $\sigcl$ becomes equal to that in the {\canonical} regime.\\
The uncertainty on the {\dpsiname} determined by the conventional method can be used at $S/N$ larger than $1$ in the {\canonical} case of the noise covariance matrix and gives a very conservative estimate of the true uncertainty.

\section{Other estimators}
\label{sec:recipes}

\subsection{Dichotomic estimator}
\label{sec:dichotomic}

\begin{figure}[!h!t]
\includegraphics[width=9cm]{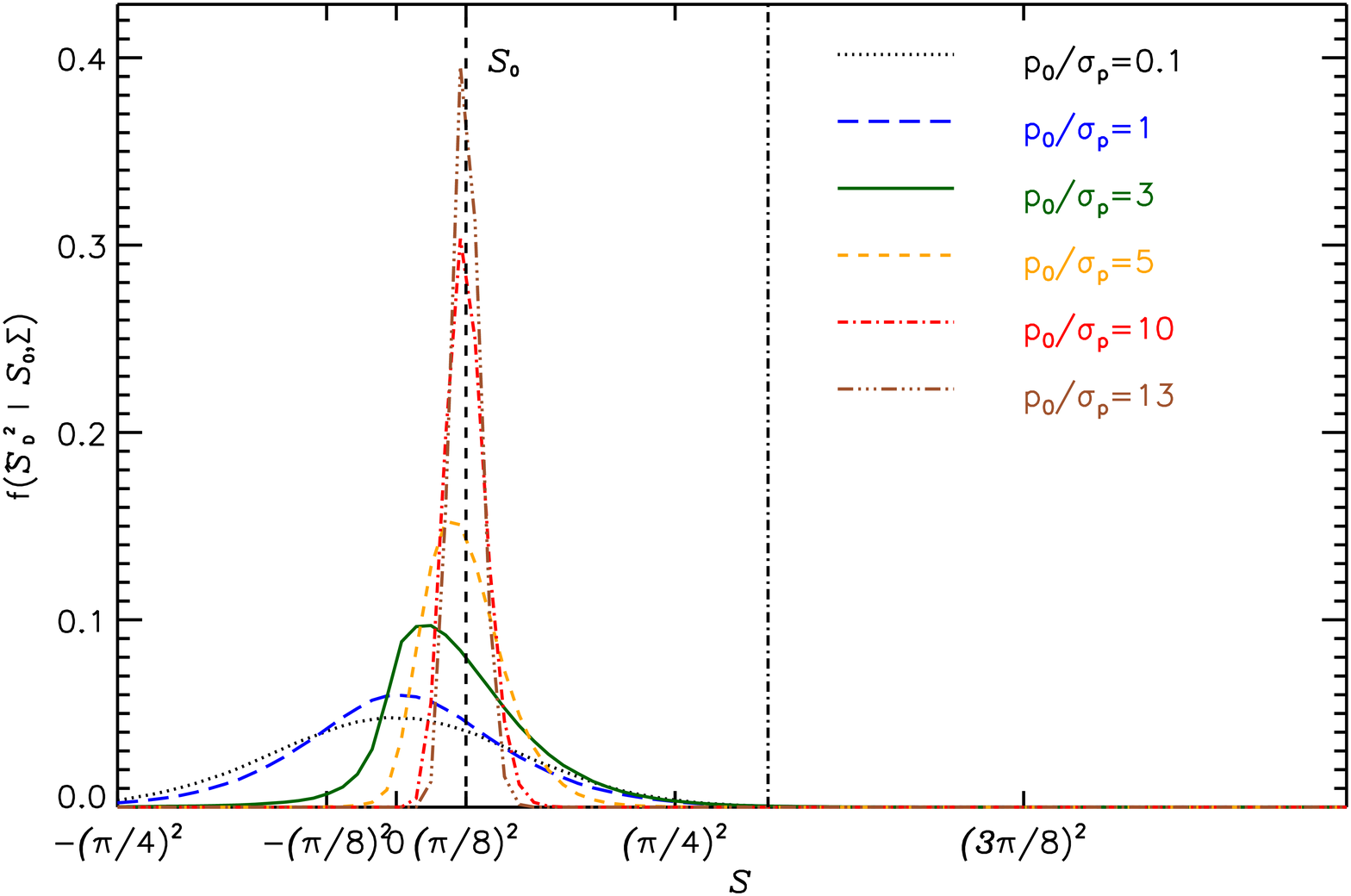}
\includegraphics[width=9cm]{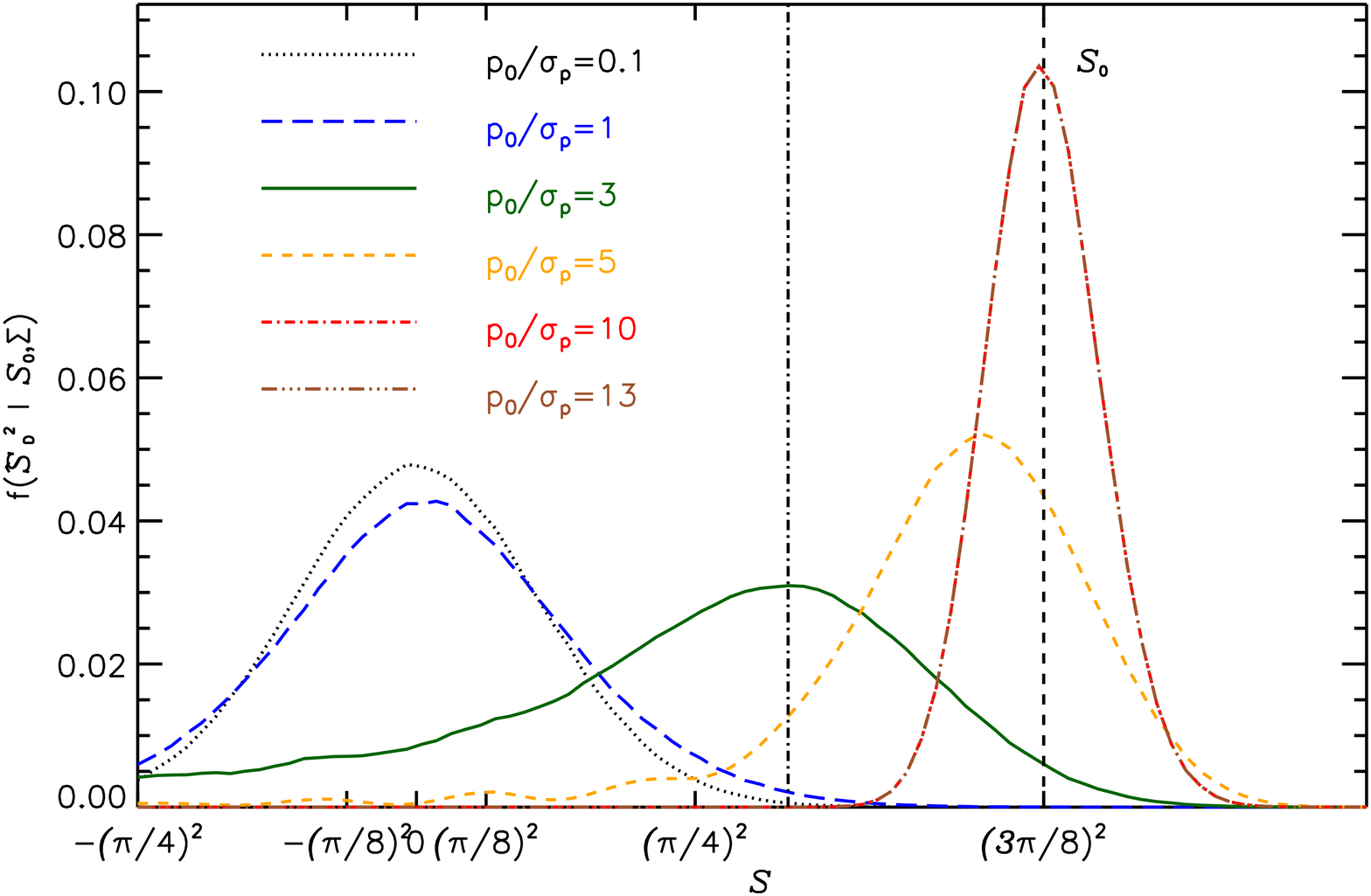}
\caption{Examples of the distribution function of the dichotomic estimator $\dich^2$ in the {\canonical} regime ($\epseff=1$). Top: $\dpsiz = \pi/8$. Bottom: $\dpsiz = 3\pi/8$. Note squared values. The vertical dashed line shows the true value and the vertical dash-dotted line shows the value of $\pi^{2}/12$.}
\label{fig:dichotomic_pdf}
\end{figure}

\begin{figure}[!h!t]
\includegraphics[width=9cm]{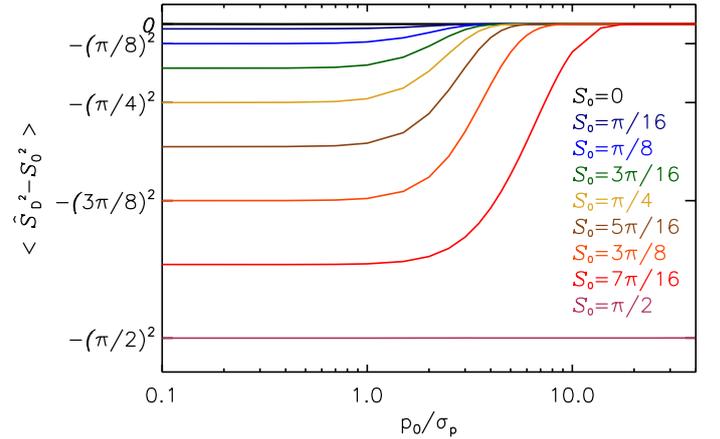}
\caption{The average bias on $10^{6}$ MC realizations of the dichotomic estimator $\dich^2$ in the {\canonical} case of the noise covariance matrix: $\epseff=1$ for the true values of $\dpsiz$ varying between $0$ and $\pi/2$ as a function of {\sn}. The colored curves are shown from top to bottom in the same order as the legend lines on the right part of the Figure.}
\label{fig:dichotomic_dpsi}
\end{figure}

The bias on the {\dpsiname} occurs because of the non-linearity in the Equation \ref{eq:dpsi_QU} when deriving $\cl$ from the Stokes parameters.
In order to overcome this issue, one can use the dichotomic estimator that consists of combining two independent measurements of the same quantity.
The square of the dichotomic estimator of the {\dpsiname} has the following form:
\begin{equation}
\hat{\dpsi^{2}_{D}} (\mathbf{x},l) =  \frac{1}{N(l)}  \sum_{i=1}^{N(l)} \left[\psi_{1}(\mathbf{x}) - \psi_{1} (\mathbf{x}+\mathbf{l}_i ) \right] \left[(\psi_{2}(\mathbf{x}) - \psi_{2} (\mathbf{x}+\mathbf{l}_i )) \right]   \, ,
\end{equation}
where subscripts $1$ and $2$ correspond respectively to each of the two data sets.
We study the behavior of the dichotomic estimator of $\dpsi^2$ by assuming the noise level of the two data sets to be $\sqrt{2}$ times lower than the noise level considered for the conventional estimator $\cl$.
This allows us to reproduce the situation where the original data had been divided in two subsets, so that $\sigp$ becomes $\sqrt{2}\,\sigp$ (as in the case of the \textit{Planck} satellite data).
The true angles are considered to be in the "uniform" configuration and the noise covariance matrix is in the {\canonical} regime.
Figure \ref{fig:dichotomic_pdf} shows the examples of the  DFs of $\dich^2$ for $\dpsiz=\pi/8$ and $\dpsiz = 3\pi/8$.
At low $S/N$s, the mean estimate of the  DFs, $<\dich^2>$ tends to $0$.
The same trend is observed for any $\dpsiz$.
The average bias for different values of $\dpsiz$ is shown in Figure \ref{fig:dichotomic_dpsi}.
We conclude that the dichotomic estimator of the {\dpsiname} is always negatively biased.

The dichotomic estimator $\dich^2$ is not suitable for accurate estimate of the {\dpsiname} because it is a quadratic function that can take negative values.
However, as its behavior is opposite to that of $\cl$ in the range $\dpsiz \, \in \, [0, \randval]$, it can be used as a verification of the validity of $\cl$:
\begin{itemize}
\item if $\cl > \randval$ and $\dich^2 > \pi^2/12$, then the noise level is low, $\dpsiz$ is larger than $\randval$, and $\cl$ gives a reliable estimate of $\dpsiz$;
\item if $\cl > \randval$ and $\dich^2 < \pi^2/12$, then the noise level is high and $\dpsiz$ is probably larger than $\randval$. In this case we suggest to estimate the upper limit of the bias as described in Section \ref{sec:maxbias}; 
\item if $\cl < \randval$ and $\dich^2 < \pi^2/12$, then $\dpsiz$ is smaller than $\randval$. We propose to use a polynomial combination of both $\cl^2$ and $\dich^2$ to better estimate $\dpsi$ (see Section \ref{sec:poly}) if two independent data sets are available, or to estimate the upper limit of the bias as described in Section \ref{sec:maxbias}.
\end{itemize}

\subsection{Bayesian DFs of $\dpsi$}
\label{method}

In an attempt to develop an accurate estimator of the {\dpsiname}, we use the difference between the behaviors of the conventional and dichotomic estimators in the range $\dpsiz \in [0, \randval]$.
In order to obtain $\dpsiz$ knowing $\cl$ and $\dich^2$ from the data, we use the Bayes' theorem.
The posterior  DF of $\dpsiz$ can be given by
\begin{equation}
 D(\dpsiz \vert \cl^{2}, \dich^{2}, \Sigma) = \dfrac {g(\cl^{2},\dich^{2} \vert \dpsiz, \Sigma) k (\dpsiz)} {\int_{0}^{\pi/2}g(\cl'^{2},\dich'^{2} \vert \dpsiz', \Sigma) k(\dpsiz') d\dpsiz'}  \, ,
 \end{equation}
 where $k (\dpsiz)$ is a prior on $\dpsiz$, which we choose to be flat in the range $[0, \pi/2]$. 
Here, $g(\cl^{2},\dich^{2} \vert \dpsiz, \Sigma)$ is the distribution function of the conventional and dichotomic estimators knowing the true {\dpsiname} $\dpsiz$ and the noise covariance matrix.

We numerically build the posterior DFs $D(\dpsiz \vert \cl^{2}, \dich^{2}, \Sigma)$ for different values of $\dpsiz$ and different {\sn} in the {\canonical} regime.
For this purpose, we first define a two-dimensional grid $G$ of the size $N_c \, \times \, N_d$ where $N_c$ and $N_d$ are the numbers of sampling of the squared conventional and dichotomic estimators in the ranges $[0,(\pi/2)^2]$ and $[-(\pi/2)^2,(\pi/2)^2]$, respectively. 
$N_c$ and $N_d$ are chosen in a way to make sure that the meshes of the grid are squares with the size of $0.00826$ $\mathrm{rad^2}$ ($N_c=300$, $N_d=600$).
Second, we run MC simulations for $\dpsiz \, \in \, [0, \, \pi/2]$ as previously. 
For each $\dpsiz$ there are $N_{MC} = 10^6$ noise realizations in the \textit{canonical} case of the noise covariance matrix, giving $N_{MC}$ pairs of $(\cl_k^2, \, \dich_k^2)_k$, where $k \, \in \, [1,\, N_{MC}]$.
After each run $k$, the corresponding $\dpsiz$ is attributed to the mesh of the grid with coordinates $(\cl_k^2, \, \dich_k^2)_k$.
Finally, we average over $\dpsiz$ in each mesh and obtain a grid of $\sbar$. 

\begin{figure*}[ht!]
\begin{center}
\includegraphics[width=15 cm]{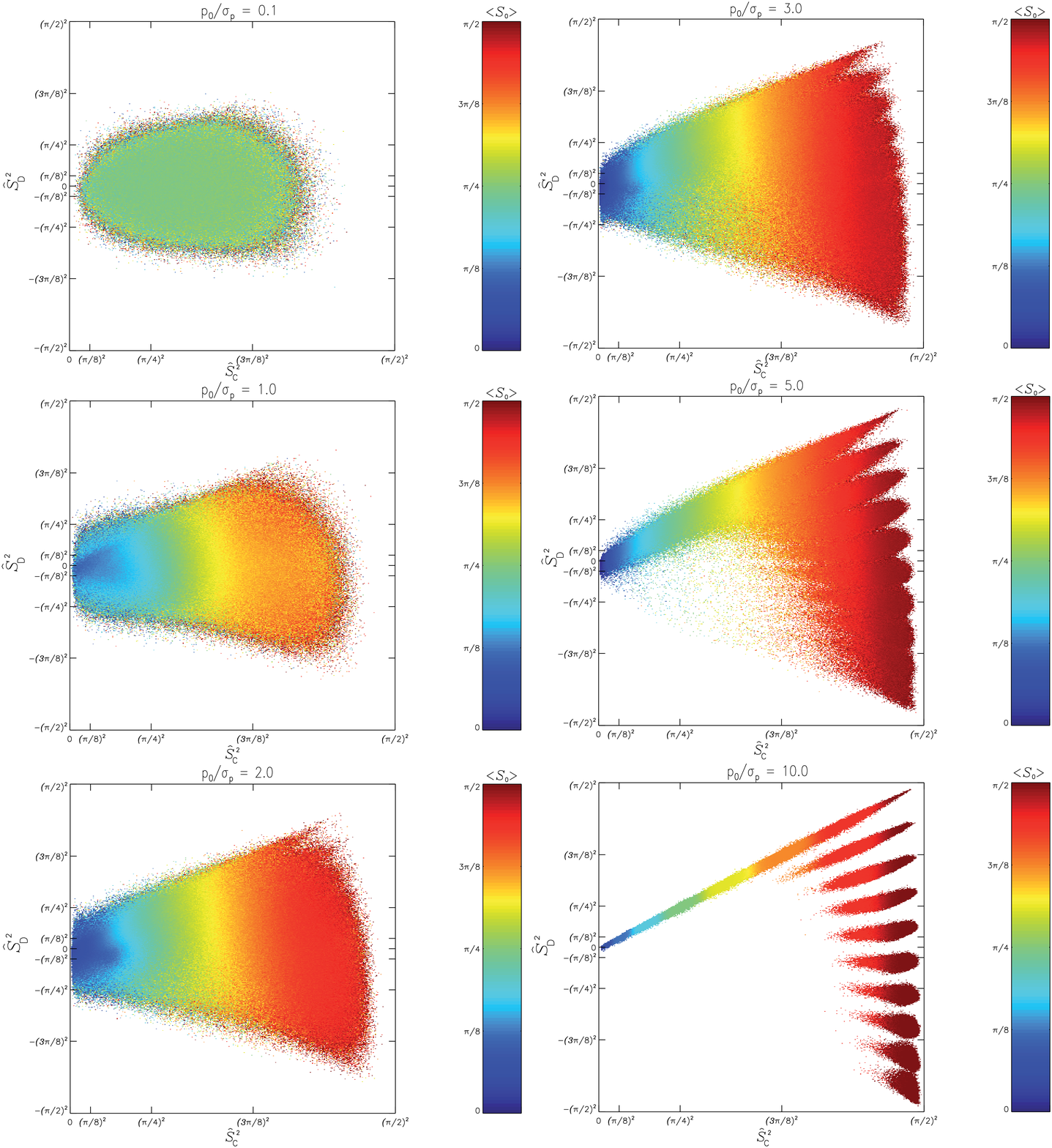}
\caption{The average of $\dpsiz$ over the posterior distribution functions of $D(\dpsiz \vert \cl^{2}, \dich^{2}, \Sigma)$ for $\snrpo \, = \, 0.1, \, 1, \, 2$ (left column, from top to bottom) and $3, \, 5, \, 10$ (right column, from top to bottom) simulated in the {\canonical} case of the noise covariance matrix.}
\label{fig:surface_result}
\end{center}
\end{figure*}

Examples of $\sbar$ for different {\sn}s in the {\canonical} case of $\Sigma_p$ are shown in Figure \ref{fig:surface_result}.
One can see that, at very low {\sn} (top left panel) almost all combinations of the two estimators give $\sbar$ distributed around $\pi/4$.
Because of the noise, both estimators fail to correctly estimate $\dpsiz$ and all possible $\dpsiz \, \in \, [0, \, \pi/2]$ give $\pi/4$ on average. 
But already at $\snrpo = 1$ (middle left panel), there is a correlation between $\cl^2$ and $\sbar$ and small variations of $\sbar$ with $\dich^2$ appear.
At intermediate {\sn} ($\snrpo =2, \, 3$, bottom left and top right panels respectively) the dependence of $\sbar$ on $\cl^2$ is the most marked: $\sbar$ is correlated with $\cl^2$ for any $\dich^2$.
In fact, as the posterior approach forces $\sbar$ to be positive, and as $\cl$ is positive by definition, this explains that $\sbar$ depends strongly on the conventional estimator.
Also, high-value $\dich^2$ are difficult to obtain at low and intermediate {\sn} as it tends to $0$ in presence of noise.
On the contrary, the dependence of $\sbar$ on the dichotomic estimator is stronger at low $\vert \dich^2 \vert$ and low $\cl^2$ (dark blue to light blue variations in panels corresponding to $\snrpo = 1,\,2,\,3$). 
At higher $\sn$ ($\snrpo > 5$, center and bottom right panels), there is a strong correlation of $\sbar$ with both $\cl^2$ and $\dich^2$. 
At these $\sn$, $\dich^2$ takes positive values for moderate $\dpsiz$, but as soon as $\dpsiz$ approaches $\pi/2$, $\dich^2$ is not efficient and we observe a feather-like pattern. 
Note that some values of $\cl^2$ and $\dich^2$ are never reached, or, in other words, there are values of $\cl$ and $\dich^2$ which do not give any $\sbar$.
We would like to emphasize that the empirical Bayesian approach used here never gives $0$ even at low ($\cl, \dich^2$) as we average over the values defined between $0$ and $\pi/2$. 

\subsection{Polynomial estimator}
\label{sec:poly}

In order to be able to directly use the conventional and dichotomic estimators of $\dpsi^2$, without computing the Bayesian Posterior  DFs, we search for a polynomial combination of $\cl^2$ and $\dich^2$ which would reflect the above simulations.
To do so, we fit the surface $\sbar$ by a polynomial of the following form:
\begin{equation}
\poly = \sum C_{a,b,n} (\cl^{2})^{a} (\dich^{2})^{b} \, ,
\end{equation}
where $a \ \in \ [0,\, n]$, $b  \ \in \ [0,\, n]$ and $n$ is the order of the polynomial.
Thus, for each {\sn} and a given order, one would have the corresponding coefficients $C_{a,b,n}$.
By applying these coefficients to any couple $(\cl^2, \, \dich^2)$ at a given {\sn}, one should be able to obtain the polynomial estimator $\poly$. 

Polynomial orders from $1$ to $6$ have been tested via comparison of the estimator $\poly$ to the result of the simulations $\sbar$.
We focused on the case of the intermediate {\sn} ($\snrpo = 2$), as it corresponds to the regime where the bias on $\cl$ is the most affected by irregularities in the shape of $\Sigma_p$.
The polynomial order $4$ is the best compromise between the order of the polynomial degree and the goodness of the fit. 

Once $C_{a,b,n}$ are known, one can apply them to any couple of the measured estimators $(\cl^{2}, \dich^{2})$ in order to calculate the polynomial estimator .  
Nonetheless, one should be cautious about unrealistic values such as low $\cl^2$ and high $\vert \dich^2 \vert$, where no correct result can exist.

\begin{figure}[hp!]
\includegraphics[width=8cm]{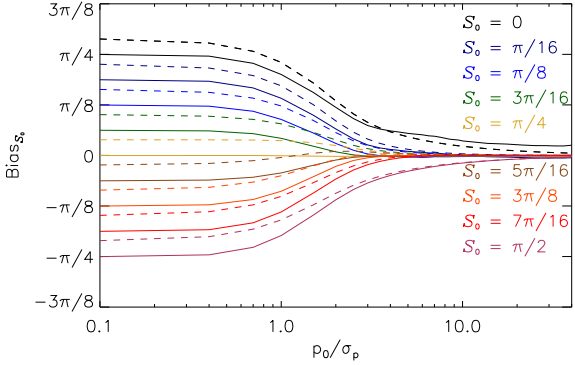}
\caption{Average bias on $10^6$ MC realizations on conventional (dashed curves) and polynomial (plain curves) estimators in the {\canonical} case of covariance matrix ($\epseff = 1$) for various $\dpsiz$ as a function of $\snrpo$. The colored curves are shown from top to bottom in the same order as the legend lines on the right part of the Figure.}
\label{fig:poly_bias}
\end{figure}

The average biases of the polynomial and conventional estimators in the {\canonical} regime and "uniform" configuration of the true angles are compared in Figure \ref{fig:poly_bias} for different {\sn}s and $\dpsiz$.
In the range $\dpsiz \in [0, \pi/\sqrt{12}]$, the conventional estimator biases positively, while the dichotomic one negatively: their contributions are opposite, and $\poly$ gives more reliable results and performs better than $\cl$ at low and intermediate {\sn}s. 
For example, at $\snrpo=2$, the bias on $\poly$ is as high as $88\%$ of the bias on $\cl$ at $\dpsiz = 0$ and it vanishes completely towards $\dpsiz=\pi/4$.
Beyond the {\sn} of $4$, the polynomial estimator is less accurate than the conventional one. 
For $\dpsiz \in [\pi/\sqrt{12}, \pi/2]$, the bias for both conventional and dichotomic estimators is negative and $\poly$ fails compared to the conventional estimator, as expected.

In this study, contributions of $\cl$ and $\dich$ have been supposed to be equal, because $\dpsiz$ is not known a priori. 
As a step forward, one can iterate on priors on $\cl$ and $\dich$ in order to improve the estimation of $\dpsiz$.
When the first approximate result is obtained and the tendency with respect to high/low $\dpsiz$ is recognized, one could attribute more or less weight to the estimator that is effective in that range of $\dpsiz$.

\subsection{Estimation of the upper limit of the bias on $\cl$}
\label{sec:maxbias}
When the dichotomic estimator cannot be calculated, i.e, there is only one measurement per spatial position, it is helpful to evaluate to which extent one can trust the conventional estimator, given by Equation \ref{eq:dpsi_QU}.
We propose a simple test that consists of calculating the maximum bias due to the noise of the data. \\
As seen in Section \ref{sec:bias}, the largest bias occurs for $\dpsiz = 0$.
A MC noise simulation consistent with the noise covariance matrices of the data at $\dpsiz=0$ would give the value of the maximum possible bias. 
For that purpose we need to change $I$, $Q$ and $U$ in such a manner as to have $\dpsiz = 0$, and we keep the {\sn} of $p$ unchanged. 
The only way to have $\dpsiz=0$ is to attribute the same true polarization angle for all the pixels inside the considered area.
Such a configuration is given by
\begin{equation}
\dfrac{\newU}{\newQ}=r \, ,
\end{equation}
where $r$ is a real constant, $\newU$ and $\newQ$ are the Stokes parameters which will be used in the calculation of the upper limit on the bias on the {\dpsiname}.
The total intensity should also be modified in order to preserve $p$.
It is given by
\begin{equation}
\newI = \dfrac{ \sqrt{\newQ^2(1+r^{2} )} } {p} \, .
\label{eq:new_q}
\end{equation}
The system for ($\newI,\newQ,\newU$) can be closed if we adopt an expression for $\sigp$.
We consider $\sigp$ as given by the conventional uncertainty estimator with no cross-correlation terms:
\begin{equation}
\sigp = \dfrac{\sqrt{Q^{2}\sigma_{Q}^{2} + U^{2}\sigma_{U}^{2} + p^{4}I^{2}\sigma_{I}^{2}}}{pI^{2}} \, .
\end{equation}
Then, the new Stokes $Q$ parameter is given by
\begin{equation}
\newQ = \dfrac{p\sqrt{\sigma_{Q}^{2} + r^{2} \sigma_{U}^{2} + p^{2}(1+r^{2}) \sigma_{I}^{2}}}{(1+r^{2})\sigp} \, ,
\end{equation}
and the expression of the new Stokes $U$ parameter is the following:
\begin{equation}
\newU = r \, \newQ \, .
\label{eq:new_u}
\end{equation}

For example, we take the true value $\dpsiz = 22.5^{\circ}$ in the uniform configuration of the true angles and the effective ellipticity $\epseff = 1.1$ ({\low} regime) with $\eps = 1.1$ and $\rho=0$. 
We assume the total intensity $I_0$ is equal to $1$ and perfectly known as in the above simulations, so that we deal with the reduced noise covariance matrix (see Equation \ref{eq:sigma_simple}).
We also assume the uncertainty $\sigma_{U}=U_0$, then $\sigma_{Q} = \eps \, \sigma_{U} = 1.1 \sigma_{U}$ from Equation \ref{eq:epsrho}.
This allows us to build the simulated noise covariance matrix $\Sigma_p$. 
We simulate a measurement by running one noise realization consistent with $\Sigma_p$ and obtain $\cl=43.1^{\circ}$.
We follow the above-described procedure and, averaging over $10^6$ noise realizations we obtain the mean value of the maximum bias $< Bias_{max} > = 21.5^{\circ}$ with the standard deviation $\sigma(Bias_{max}) = 7.5^{\circ}$.
Thus, in this case, the estimation of $\dpsi$ can be affected by bias almost by the same order of magnitude as the true value.
This method can not be directly used to "de-bias" the conventional estimator but can be used to estimate, on average, at which level the estimation of the {\dpsiname} is affected by the noise level and the shape of the noise covariance matrix.

\section{Discussion and conclusion}
\label{sec_conclusion}

In this paper, we studied the bias on the {\dpsiname} and we have demonstrated its complex behavior for the first time.
We showed that it strongly depends on the true value which is not known a priori: the bias on the conventional estimator is negative for $\dpsiz > \randval$ ($\simeq 52^{\circ}$), which is the value corresponding to the result if all the angles considered in the calculation are random, positive for $\dpsiz < \randval$, and it can reach up to $\randval$ at low {\sn}s (Section \ref{sec:bias_dpsiz}).
The bias on the {\dpsiname} also depends on the shape of the noise covariance matrix and the distribution of the true angles in the intermediate range of {\sn}, between $1$ and $4$ as seen in Sections \ref{sec:cov_matrix}, \ref{sec:true_angle_eps1}.
However, if there is less than $10\, \%$ effective ellipticity between noise levels on Stokes parameters $Q$ and $U$, the impact of the shape of the noise covariance matrix and of the distribution of the true angles can be neglected.
Otherwise, these factors can significantly affect the estimation of the {\dpsiname} when using the conventional estimator.

We have introduced the dichotomic estimator of $\dpsi$ and studied its behavior. 
We showed that the bias on $\dich^2$ is always negative. 
In addition, such an estimator has the disadvantage of being a quadratic function that can take negative values.
However, using both conventional and dichotomic estimators appears to be the first step in assessing the true value of the {\dpsiname}.
We have introduced a new polynomial estimator that allows us to use the low {\sn} data (less than $4$).
This broadens the application of the {\dpsiname} in different polarimetric studies. 
Yet deriving the polynomial estimator requires the existence of at least two independent measurements as well as an additional computational time to run simulations.

We propose a method to evaluate the maximum possible bias of the {\dpsiname} knowing the noise covariance matrix of the data.
It can be used as an estimator of the upper limit to the bias on $\cl$ with any polarimetric data with the available noise covariance matrices in $(Q,U)$.

The methods developed in this work (maximum bias estimation and dichotomic estimator) have been applied to the \textit{Planck} data in order to analyze the observed dust polarization with respect to the magnetic field structure. 
\cite{planck2014-xix} calculated the {\dpsiname} in an annulus of a $30'$ lag and $30'$ width all over the sky at $1^{\circ}$ resolution, revealing filamentary features.
Using the dichotomic estimator and the test of the maximum bias on $\dpsi$, \cite{planck2014-xix} demonstrated that these filamentary features are not artifacts of noise.
Moreover, a clear anti-correlation between the polarization fraction and the {\dpsiname} has been shown. \\
\cite{planck2014-xix} used the data smoothed to $1^{\circ}$ resolution, which diminishes the noise level.
Also, as the effective ellipticity of the \textit{Planck} data deviates at most by $12\, \%$ from the {\canonical} case \citep{planck2014-xix}, the shape of the noise covariance matrix has been taken into account in the estimation of $\dpsi$. 
The results of this work can also be particularly well suited in the analysis of the data from the new experiments that are designed for polarized emission studies, such as the balloon-borne experiments BLAST-Pol \citep{fissel2010}, PILOT \citep{bernard2007} and the ground-based telescopes with new polarization capabilities: ALMA \citep{perez-sanchez2013}, SMA, NIKA2 \citep{catalano2016}.
We suggest to calculate both the conventional and dichotomic estimators in order to compare both, in the case where two independent data-sets are available, as well as to estimate the upper limit of the bias on $\dpsi$ using the method proposed in this work for any polarimetric data with the noise covariance matrix provided.
A joint IDL/Python library which includes the methods from the work on bias analysis and estimators of polarization parameters is currently under development.



\appendix


\section{Derivation of the conventional uncertainty}
\label{app_derivatives}

We assume the uncertainties on angles to be known.
Let start by the definition of variance applied to $\dpsi$ and consider small displacement of $\dpsi$:
\begin{equation}
\sigma^{2}_{\dpsi (\mathbf{x},l) }  = E [ (\dpsi (\mathbf{x},l) - E [ \dpsi (\mathbf{x},l) ] )^{2}] = E [ (d \dpsi (\mathbf{x},l))^{2} ] \, .
\end{equation}
The differential of $\dpsi$ includes partial derivatives with respect to the angle at position $\mx$ and each angle at positions $\mathbf{x}+\mathbf{l_i}, \ \mbox{with} \ i \ \in [1,\,N]$:
\begin{equation}
d\dpsi (\mathbf{x},l ) =\dfrac {\partial \dpsi (\mathbf{x},\mathbf{l}) } {\partial \psi (\mathbf{x}) } d \psi (\mathbf{x}) + \sum_{i=1}^{N} \Big[\dfrac { \partial \dpsi }{\partial \psi (\mathbf{x}+\mathbf{l_{i}} ) } d \psi (\mathbf{x}+\mathbf{l_{i}} )\Big] \, .
\end{equation}
When developing the square, one has:
\begin{align}
( d\dpsi (\mathbf{x},l ) ) ^ {2} \ = \ & ( \dfrac {\partial \dpsi (\mathbf{x},l) } {\partial \psi (\mathbf{x}) } ) ^{2} \,( d \psi (\mathbf{x}) ) ^ {2} \nonumber \\
&+  \sum_{i=1}^{N} \, ( \dfrac { \partial \dpsi (\mathbf{x},l )}{\partial \psi (\mathbf{x}+\mathbf{l_{i}} ) } ) ^ {2} \, ( d \psi (\mathbf{x}+\mathbf{l_{i}} ) ) ^{2}  \nonumber \\
&+ 2 \sum_{i=1}^{N} \, \dfrac {\partial \dpsi (\mathbf{x},l) } {\partial \psi (\mathbf{x}) }  \dfrac { \partial \dpsi (\mathbf{x},l )}{\partial \psi (\mathbf{x}+\mathbf{l_{i}} ) } \, d \psi (\mathbf{x}) \, d \psi (\mathbf{x} + \mathbf{l_{i}})  \, .
\label {eq:dpsi_sig_full}
\end{align}
If one takes the expectation of $d\dpsi^2$, then
\begin{align}
 E [d\dpsi (\mathbf{x},l ) ^ {2} ] \ = \ &  ( \dfrac {\partial \dpsi (\mathbf{x},l) } {\partial \psi (\mathbf{x}) } ) ^{2}  \, \sigma^{2}_{ \psi (\mathbf{x})}  + \sum_{i=1}^{N} \, ( \dfrac { \partial \dpsi (\mathbf{x}, l )}{\partial \psi (\mathbf{x}+\mathbf{l_{i}} ) } ) ^ {2}   \,\sigma^{2}_{\psi (\mathbf{x+l_{i}})}  \nonumber \\
+ & 2 \, \sum_{i=1}^{N} \, \dfrac {\partial \dpsi (\mathbf{x},l) } {\partial \psi (\mathbf{x}) }   \dfrac { \partial \dpsi (\mathbf{x}, l )}{\partial \psi (\mathbf{x}+\mathbf{l_{i}} ) }  \, \sigma_{\psi (\mathbf{x}) \psi (\mathbf{x+l_{i}}) } \, .
\label {eq:dpsi_sig_full2}
\end{align}

The partial derivatives are:
{\small 
\begin{equation*}
\dfrac{\partial \dpsi (\mathbf{x},l )  } {\partial \psi (\mathbf{x}) } = \dfrac{1}{2} \, \Big( \dfrac{1}{N} \, \sum_{i=1}^{N} \,[\psi (\mathbf{x}) -\psi (\mathbf{x+l_{i}}) ]^{2} \Big)^{-1/2 } \, \Big( \dfrac{2}{N}\, \sum_{i=1}^{N} \, [\psi (\mathbf{x}) - \psi (\mathbf{ x+l_{i} } ] \Big) \, , 
\end{equation*}
\begin{equation*}
\dfrac {\partial \dpsi (\mathbf{x},l ) } {\partial \psi (\mathbf{x+l_{i}}) } = - \dfrac{1}{2} \, \Big( \dfrac{1}{N} \, \sum_{i=1}^{N} \, [ \psi (\mathbf{x}) -  \psi (\mathbf{x} +\mathbf{l}_{i} )]^{2} \Big)^{-1/2 } \, \dfrac{2}{N} \, \Big(\psi (\mathbf{x}, \mathbf{l} ) - \psi (\mathbf{x} + \mathbf{l}_{i} )\Big) \, ;
\end{equation*}
\begin{eqnarray}
\Big(\dfrac{\partial \dpsi (\mathbf{x},l )  } {\partial \psi (\mathbf{x}) } \Big) ^{2} & = & \dfrac{1}{N^{2}} \Big( \dfrac{1}{N}  \sum_{i=1}^{N} [  \psi (\mathbf{x} ) - \psi (\mathbf{x} +\mathbf{l}_{i} ) ]^{2} \Big)^{-1 } \nonumber \\ &&  \quad \times \,
 \Big(\sum_{i=1}^{N}[ \psi (\mathbf{x}) -  \psi (\mathbf{x} +\mathbf{l}_{i} ) ]\Big)^{2} \nonumber \\
& = & \dfrac { \Big(\sum_{i=1}^{N}[ \psi (\mathbf{x}) -  \psi (\mathbf{x} +\mathbf{l}_{i} ) ] \Big)^{2} } { N^{2} [\dpsi (\mathbf{x},l) ]^{2}} \\
\Big(\dfrac {\partial \dpsi (\mathbf{x},l ) } {\partial \psi (\mathbf{x+l_{i}}) }  \Big) ^ {2} & = & \dfrac{ [ \psi (\mathbf{x} ) - \psi (\mathbf{x} +\mathbf{l}_{i} ) ]^{2} }{N^{2}[\dpsi (\mathbf{x},l) ]^{2}} 
\end{eqnarray}
}
As the noise levels on two measurements of polarization angle at different positions are uncorrelated, one has:
\begin{equation*}
\sigma_{\psi (\mathbf{x}) \psi (\mathbf{x} + \mathbf{l_{i}}) } = 0 \, .
\end{equation*}
Since $E [d\dpsi (\mathbf{x},l ) ) ^ {2} ] = \sigma^{2}_{\dpsi (\mathbf{x},l)}$, Equation \ref{eq:dpsi_sig_full} becomes
\begin{eqnarray}
\sigma^{2}_{\dpsi (\mathbf{x},l)} =  && \dfrac{1}{[N\dpsi (\mathbf{x},l) ]^{2}} \Big[ \Big(\sum_{i=1}^{N}[ \psi (\mathbf{x}) -  \psi (\mathbf{x} +\mathbf{l}_{i} ) ] \Big)^{2} \sigma^{2}_{\psi(x)}  \nonumber \\
& & \quad + \sum_{i=1}^{N} (\psi (\mathbf{x}) -  \psi (\mathbf{x} +\mathbf{l}_{i} ) )^{2} \sigma^{2}_{\psi(\mathbf{x+l_{i}})} \Big] \, .
\end{eqnarray}
Taking the square root of this expression, one gets the conventional uncertainty on {\dpsiname}:
\begin{eqnarray}
\sigcl  =  \dfrac {1}{N \dpsi (\mathbf{x},l)} & & \Big[\Big(\sum_{i=1}^{N}[ \psi (\mathbf{x}) -  \psi (\mathbf{x} +\mathbf{l}_{i} ) ]\Big)^{2} \sigma^{2}_{\psi(\mathbf{x})} \nonumber \\
& &+ \sum_{i=1}^{N} [\psi (\mathbf{x}) -  \psi (\mathbf{x} +\mathbf{l}_{i} ) ]^{2} \sigma^{2}_{\psi(\mathbf{x+l_{i}})}\Big]^{1/2} \, .
\end{eqnarray}

\end{document}